\documentstyle[epsf]{article}
\def\vec#1{\mathchoice
{\mbox{\boldmath $\displaystyle#1$}}
{\mbox{\boldmath $\textstyle#1$}}
{\mbox{\boldmath $\scriptstyle#1$}}
{\mbox{\boldmath $\scriptstyle#1$}}}

\newcommand{\ket}[1]{| #1 \rangle}

\newcommand{\experiment}[1]{\langle #1 \rangle}
\newcommand{\element}[3]{\langle #1 | #2 | #3 \rangle}
\newcommand{\Cr}[1]{C^{\rho}_{#1}}
\newcommand{\Cdr}[1]{C^{\triangle \rho}_{#1}}
\newcommand{\Cta}[1]{C^{\tau}_{#1}}
\newcommand{\CJ}[1]{C^{J}_{#1}}
\newcommand{\CdJ}[1]{C^{\nabla J}_{#1}}
\newcommand{\Cs}[1]{C^{s}_{#1}}
\newcommand{\Cds}[1]{C^{\triangle s}_{#1}}
\newcommand{\nn}{\nonumber}

\newcommand{\al}{\alpha}
\newcommand{\be}{\beta}
\newcommand{\ga}{\gamma}
\newcommand{\de}{\delta}
\newcommand{\si}{\sigma}
\newcommand{\da}{\dagger}

\begin{document}
\title{Oblate-prolate transition in odd-mass light mercury isotopes}
\author{
Shouichi Sakakihara\footnote{
{\em e-mail address \rm: shouichi@npl.kyy.nitech.ac.jp}}
\ and Yasutoshi Tanaka
\\
{\em Department of  Environmental Technology and Urban Planning}, \\
{\em Nagoya Institute of Technology, Gokiso, Nagoya 466-8555, Japan}\\
}
\maketitle
\begin{abstract}
Anomalous isotope shifts in the chain of light Hg isotopes are 
investigated by using the Hartree-Fock-Bogoliubov method 
with the Skyrme SIII, SkI3 and SLy4 forces. 
The sharp increase in the mean-square radius of 
the odd mass $^{181-185}$Hg isotopes is 
well explained in terms of the transition 
from an oblate to a prolate shape 
in the ground state of these isotopes. 
We discuss the polarization energy of time-odd mean-field
terms in relation to the blocked level by the odd neutron.
\end{abstract}

\section{Introduction}

The sharp increase in the mean-square charge radius is 
observed in the odd mass $^{181-185}$Hg isotopes 
with respect to their even neighbors \cite{AH87}. 
A hint that a very different configuration may constitute 
the ground state of these nuclei was given 
by an isomer shift measurement 
of $^{185}$Hg \cite{DB79}, 
where they found an excited state of $^{185}$Hg 
which has a charge radius consistent with
the charge radius of heavier neighbors. 
The sharp increase in the charge radius was 
interpreted as the transition from 
a nearly spherical to a deformed shape 
in the ground state of the odd mass $^{181-185}$Hg isotopes.

Frauendorf and Pashkevich \cite{FP75} calculated 
the deformation energy of Hg isotopes 
by means of Strutinsky's shell correction method. 
They found that the light Hg isotope 
has an oblate and prolate minimum of nearly 
the same depth at the respective 
deformations $\epsilon = -0.12$ and 0.22.  
They discussed a mechanism which causes 
the transition from a small oblate 
to a large prolate deformation in the odd mass $^{181-185}$Hg isotopes. 
They noticed that the neutron level density 
is very high in the almost spherical oblate minimum, 
whereas the neutron level density is low in 
the well deformed prolate minimum. 
They concluded that the loss of pairing energy 
due to the blocking of the Fermi level 
by the odd neutron is larger in the oblate minimum 
than in the prolate one, which gives rise to 
the oblate-prolate transition in the ground state 
of these isotopes. 
As the prolate minimum corresponds to deformation 
twice as large as the oblate one, 
the sharp increase in the mean-square radius 
is observed in the odd-mass light Hg isotopes.

There are several mean field calculations 
of Hg isotopes. 
However, all of them are limited 
to even isotopes \cite{CL73,YT97,RR00,NV02}. 
In order to see if the mechanism of  
Frauendorf and Pashkevich works well, 
we perform Hartree-Fock-Bogoliubov (HFB) calculations 
of Hg isotopes with effective interactions 
SIII \cite{BF75}, SkI3 \cite{RF95} and SLy4 \cite{CB98}. 
We calculate Lipkin-Nogami corrections \cite{LI60,NO64} 
to both mean field and 
pairing energies \cite{BR00,SW94}. 
We investigate the influence of 
the time-odd mean-field terms as well as the blocking of 
a level by the odd neutron on the oblate-prolate shape transition.

In Section 2, we present 
the Skyrme HFB method with Lipkin-Nogami corrections 
to describe both even and odd Hg isotopes. 
In Section 3, we present calculations 
with effective interactions SIII, SkI3 and SLy4.
Conclusions are given in Section 4.

\section{Method of calculation}

\subsection{Skyrme Hartree-Fock-Bogoliubov method}

The HFB equation is derived 
by the variation of energy 
with respect to the quasi-particle vacuum \cite{RS80}, 
\begin{equation}
H=\sum_{\al\be}t_{\al \be} {c_\al}^\da {c_\be}
+\frac{1}{4}\sum_{\al\be\ga\de}\bar{v}_{\al\be\ga\de}
{c_\al}^\da {c_\be}^\da c_\de c_\ga ,
\end{equation}
\begin{equation}
\delta\frac{\element{\Phi}{H}{\Phi}}
           {\langle\Phi|\Phi\rangle}=0 .
\end{equation}
Here the quasi-particle vacuum $|\Phi\rangle$ is given by
\begin{equation}
\beta_k = \sum_{\al} 
U_{\al k}^{*} c_{\al} + V_{\al k}^{*}{c_{\al}}^{\da} ,
\end{equation}
\begin{equation}
\be_k \ket{\Phi} = 0 
\label{qp-vacumm}
\end{equation}
for all quasi-particle states $k$. 
In the present study, we employ 
Skyrme forces in the particle-hole channel 
and the zero-range density-dependent pairing force 
in the particle-particle channel,  
\begin{equation}
V_p = \frac{V_0}{2}
\left( 1 - P_{\si} \right)
\left( 1 - \frac{\rho(\vec{r_1})}{\rho_{\rm c}} \right)
\delta(\vec{r_1} - \vec{r_2}) .
\label{DDDI}
\end{equation}
The HFB energy is then given by
\begin{equation}
E^{\rm HFB} = \int {\cal H} d\vec{r} + 
          \frac{1}{2} \int \Delta(\vec{r}) \kappa^*(\vec{r}) d\vec{r} ,
\label{EHFB}
\end{equation}
where ${\cal H}$ is the Skyrme energy density \cite{VB72,EB75}
and $\Delta(\vec{r})$ is the gap field,
\begin{equation}
\Delta(\vec{r}) = \frac{V_0}{2} 
                  \left( 1 - \frac{\rho(\vec{r})}{\rho_{\rm c}} \right)
                  \kappa(\vec{r}) .
\label{Gap field}
\end{equation}
The pairing strength $V_0$ is determined from 
the odd-even mass difference and the density parameter 
$\rho_{\rm c}$ is taken as the saturation density of 
symmetric nuclear matter which is obtained
by using the respective Skyrme parameters.
The particle density $\rho(\vec{r})$ and 
the anomalous density $\kappa(\vec{r})$ are written as
\begin{equation}
\rho(\vec{r} \si, \vec{r} \si')
= \sum_{\al \be} \rho_{\al \be} 
\varphi_{\al}(\vec{r}, \si) \varphi_{\be}^* (\vec{r}, \si') ,
\label{density}
\end{equation}
\begin{equation}
\kappa(\vec{r} \si, \vec{r} \si')
= \sum_{\al \be} \kappa_{\al \be}
\varphi_{\al}(\vec{r}, \si) \varphi_{\be} (\vec{r}, \si')
\label{anomalous density}
\end{equation}
with the density matrix and pairing tensor, 
\begin{equation}
\rho_{\al \be} 
= \element{\Phi}{{c_{\be}}^{\da}c_{\al}}{\Phi}
 = \sum_k V^*_{\al k} V_{\be k} ,
\label{density matrix}
\end{equation}
\begin{equation}
\kappa_{\al \be} 
= \element{\Phi}{{c_{\be}} c_{\al}}{\Phi}
 = \sum_k V^*_{\al k} U_{\be k} .
\label{pairing tensor}
\end{equation}
Variation of Eq. (\ref{EHFB}) with respect to $\rho_{\al \be}$ and
$\kappa_{\al \be}$ yields the HFB equation,
\begin{equation}
\left(
\begin{array}{cc}
h_{\al \be} - \lambda \delta_{\al \be} & \Delta_{\al \be}  \\
-\Delta^*_{\al \be}   & -h^*_{\al \be} + \lambda \delta_{\al \be}
\end{array}
\right)
\left(
\begin{array}{c}
 U_{\be k} \\ V_{\be k}
\end{array}
\right) = E_k
\left(
\begin{array}{c}
 U_{\al k} \\ V_{\al k}
\end{array}
\right) .
\label{HFBeq}
\end{equation}
Diagonalization is carried out in a deformed oscillator basis 
with axial symmetry \cite{VA73}. 
A Greek index in Eq.~(\ref{HFBeq}) denotes 
a set of quantum numbers \{$n_r n_z \Lambda \Sigma$\} 
of a deformed oscillator wave function.

In the present study we will restrict our basis space as
\begin{equation}
E_\alpha = (2 n_r + |\Lambda| + 1) 
\hbar \omega_\perp +\left(n_z + \frac{1}{2}\right)\hbar \omega_z 
\leq\left(N_0 + \frac{3}{2}\right) \hbar \omega_0 
\end{equation}
with $N_0=18$. The oscillator constant is held fixed as 
$\hbar \omega_0 = \hbar (\omega_{\perp}^2 \omega_z)^{1/3} 
= 1.21 \times \frac{\hbar^2}{m} A^{-1/3}$. 
The deformation parameter $ q = \omega_{\perp} / \omega_z $ 
is varied to obtain the maximum deformation energy 
in the oblate and prolate minimum, respectively. 
The chemical potential $\lambda$
is adjusted by the particle number condition, 
$\experiment{\hat{N}} = N$. 
Using the occupation amplitude 
$N_k = \sum_{\al}  V^*_{\al k} V_{\al k}$ and 
quasi-particle energy $E_k$, 
we calculate an auxiliary single-particle energy and pairing gap
in order to find the optimum value of $\lambda$  \cite{DF84}, 
\begin{equation}
\bar{e}_k = - E_k (2 N_k -1 ) + \lambda ,
\label{etilde}
\end{equation}
\begin{equation}
\bar{\Delta}_k = 2 E_k \sqrt{N_k(1 - N_k)} .
\label{Dtilde}
\end{equation}

As in Ref.~\cite{SD00,DN01}, the cut-off procedure of 
quasi-particle states is imposed on 
the auxiliary single-particle energy. 
The value of  $\bar{e}_{\rm max} = 60$ MeV is 
assumed in the present study.

\subsection{Description of an odd nucleus}

The above formalism can be straightforwardly 
extended to the case of one quasi-particle state in odd mass nuclei. 
All the equation retain their form. 
Only the density matrix $\rho_{\alpha\beta}$ and 
pairing tensor $\kappa_{\alpha\beta}$ should be 
modified due to the blocked quasi-particle 
$\be_{k}^{\da} \ket{\Phi}$ in 
the following way \cite{SU66,RB70};
\begin{eqnarray}
\tilde{\rho}_{\al \be} &=&
\rho_{\al \be} + U_{\al k} U^*_{\be k} - V^*_{\al k} V_{\be k} ,
\label{density matrix2}
\\
\tilde{\kappa}_{\al \be} &=&
\kappa_{\al \be} + U_{\al k} V^*_{\be k} - V^*_{\al k} U_{\be k} .
\label{pairing tensor2}
\end{eqnarray}

We introduce a reference state $\ket{\Phi_{\rm HFBE}}$ 
and its energy expectation value $E^{\rm HFBE}$. 
The state is constructed as an {\it even} vacuum 
without quasi-particle creation and without 
breaking time-reversal invariance 
but with an {\it odd} average particle number. 
The difference of energy $E^{\rm HFB}-E^{\rm HFBE}$ 
is thus composed of the loss of pairing energy 
due to the blocking of the Fermi level 
and the polarization energy of 
the mean field by the presence of an odd neutron \cite{DB01}.

A proper description of an odd nucleus 
by mean field theories requires to break 
the time-reversal symmetry \cite{EB75}. 
When the time-reversal symmetry is broken, 
the Skyrme energy density ${\cal H}$ is written by 
the sum of kinetic energy and 
potential energy of isoscalar ${\cal H}_0$ 
and isovector ${\cal H}_1$ terms;
\begin{eqnarray}
\nn
{\cal H} &=& \frac{\hbar^2}{2m} \tau_0 + {\cal H}_0 + {\cal H}_1 ,
\\ \nn
{\cal H}_t &=&  \Cr{t}  \rho^2_t 
              + \Cs{t}  \vec{s}^2_t
              + \Cdr{t} \rho_t \triangle \rho_t
              + \Cds{t} \vec{s}_t \cdot \triangle \vec{s}_t 
\\
           && + \Cta{t} (\rho_t \tau_t - \vec{j}^2_t)
              + \CJ{t}  (\vec{J}^2_t - \vec{s}_t \cdot \vec{T}_t)
              + \CdJ{t} [  \rho_t \nabla \cdot \vec{J}_t
                         + \vec{s}_t \cdot (\nabla \times \vec{j}_t)] .
\label{functional}
\end{eqnarray}
The coupling constants $C$ are expressed by 
the Skyrme force parameters \cite{DD95}. 
The isospin index $t$ can have values 0 or 1. 
The isoscalar and isovector parts of the particle densities
$\rho_t$ are defined as 
\begin{equation}
\rho_0 = \rho_n + \rho_p , \qquad  \rho_1 = \rho_n - \rho_p ,
\end{equation}
and analogous expressions are used to define other densities. 
It should be noted that since we consider 
the Skyrme force as the density-dependent two-body interaction, 
the time-odd terms $C^s_t$ in Eq.~(\ref{functional})
are different from the original expressions \cite{EB75} 
which were derived from the three-body interaction.

The characteristic combinations of 
$(\rho_t\tau_t-\vec{j}^2_t)$, 
$(\vec{J}^2_t-\vec{s}_t\cdot\vec{T}_t)$ and 
$[\rho_t\nabla\cdot\vec{J}_t+ \vec{s}_t\cdot(\nabla\times\vec{j}_t)]$ 
in Eq.~(\ref{functional}) are the results of 
the local gauge invariance of the Skyrme force \cite{DD95}. 
The time-odd coupling constants are uniquely determined 
from the original Skyrme-force parameters and therefore 
the time-odd mean field has no adjustable parameters 
once the time-even mean field is given. 
However, as discussed by Dobaczewski and Dudek \cite{DD95}, 
one may consider the energy density to 
be a more fundamental construction than 
the Skyrme interaction itself. 
In such a case, all 20 coupling constants 
in Eq.~(\ref{functional}) can be treated 
and adjusted independently. 
However, the gauge-invariant conditions 
restrict the values of six time-odd coupling 
constants and leave the freedom to modify 
the values of $C^s_t$ and $C^{\Delta s}_t$. 
We are going to vary $C^s_t$ and $C^{\Delta s}_t$ 
to see the influence of these terms on 
the oblate-prolate transition in the odd-mass $^{181-185}$Hg
isotopes. 
We will omit the terms
$(\vec{J}^2_t-\vec{s}_t\cdot\vec{T}_t)$
because $\vec{J}^2_t$ are 
omitted in the parameterization of the present Skyrme forces.

\subsection{Lipkin-Nogami corrections}

In light Hg isotopes, 
the neutron level density is high in the oblate minimum, 
while it is low in the prolate minimum. 
Because of the difference of level densities
in the respective minima, 
it happens that neutrons 
in the oblate minimum are superconductive, 
whereas they are normal in the prolate minimum
within the same isotope. 
The situation is extremely inconvenient 
because we want discuss the energy difference between 
the oblate and prolate minimum. 
Besides, a sudden phase-transition from 
the normal to superconductive state or 
vice versa at certain neutron numbers 
may be unphysical and it may be due to 
the defect of the number non-conserving pairing scheme. 
In the present study we employ 
the approximate number-projection method of 
Lipkin and Nogami (LN) \cite{LI60,NO64} and 
calculate its corrections to both mean field
and pairing energies \cite{BR00}. 
The number projected energy is expressed as
\begin{equation}
E_{\rm HFBLN} = E_{\rm HFB} 
                     + \lambda_1 ( N - \experiment{ \hat{N} } )
                     + \lambda_2 ( N^2 - \experiment{ \hat{N}^2 } )
                     + \cdots ,
\label{EHFBLN}
\end{equation}
where $\lambda_1$ and $\lambda_2$ are 
constant parameters \cite{SW94}. 
The second order expansion is considered 
in the Lipkin-Nogami method. 
Variation with respect to the density matrix 
and pairing tensor as in the case of 
the HFB equation yields the HFBLN equation 
applicable to both even and odd nuclei
(hereafter simply called HFB). 
The parameter $\lambda_1$ is given 
by the particle number condition, 
whereas the parameter $\lambda_2$ is 
held fixed during the variation and 
is determined after the variation from the additional condition, 
\begin{equation}
\ket{\xi} = \exp (i \xi \hat{N}_{20}) \ket{\Phi} , 
\label{ss}
\end{equation}
\begin{equation}
\lambda_2 = \frac{\left.
                  \partial^2_{\xi} \element{\Phi}{H}{\xi}
                  \right|_{\xi = 0}}
                 {\left.
                  \partial^2_{\xi} \element{\Phi}{\hat{N}^2}{\xi}
                  \right|_{\xi = 0}} .
{\label{lam2}}
\end{equation}
Equation (\ref{lam2}) is calculated by the following 
density matrix and pairing tensor \cite{BR00};
\begin{eqnarray}
 \rho_{\alpha \beta}^{(\xi)} 
 &=& \element{\Phi}{ c_{\beta}^{\da} c_{\alpha} }{\xi}
  = \left( \rho + 2 i \xi (1 - \rho) \rho \right) _{\alpha \beta} ,
 \\
 \kappa_{\alpha \beta}^{(\xi)} 
 &=& \element{\Phi}{ c_{\beta} c_{\alpha} }{\xi}
  = \left( \kappa + 2 i \xi (1 - \rho) \kappa \right)_{\alpha \beta} ,
 \\
 \bar{\kappa}_{\alpha \beta}^{(\xi)}
 &=& \element{\Phi}{ c_{\alpha}^{\da} c_{\beta}^{\da} }{\xi}
  = ( \kappa^* - 2 i \xi \kappa^* \rho )_{\alpha \beta} .
\end{eqnarray}  
For an odd nucleus, the density matrix and pairing tensor 
of Eqs.~(\ref{density matrix2}) and (\ref{pairing tensor2})
are used on the right hand side of the above equations \cite{SW94}.

\section{Results and discussion}

We have analysed the oblate-prolate transition 
in the odd-mass light Hg isotopes using 
the Skyrme Hartree-Fock-Bogoliubov method 
with effective interactions SIII, SkI3 and SLy4. 
Shape transition occurs only in the 
odd-mass isotopes as the result of a subtle balance 
between the time-even mean-field energy, 
the loss of pairing energy by the odd neutron 
and the time-odd mean-field energy. 
We have examined effective interactions
SIII, SGII \cite{GS81}, SkM$^*$ \cite{BQ82}, SKX \cite{AB98}, 
SkI1, SkI2, SkI3, SkI4 and SkI5 \cite{RF95} and SLy4 
whether they can predict the oblate-prolate shape coexistence 
in light Hg isotopes. 
We have found that both SIII and SkI3 
interactions have desirable 
isospin dependence and deformation energies.

We have assumed the zero-range 
density-dependent pairing interaction.
The pairing strength $V_0$ was adjusted to reproduce 
the one-particle separation energy \cite{AW95}. 
The density parameter $\rho_c$ is taken 
as the saturation density of symmetric 
nuclear matter which is obtained from Skyrme parameters. 
Adjusted values are given in the respective figure captions.

We have investigated the effect of 
time-odd mean-field terms on the HFB energy, 
first by using the time-odd coupling constants 
derived from the Skyrme force parameters and 
second by switching on and off 
the coupling constants $C^{\Delta s}_t$ and $C^s_t$, 
respectively. These coupling constants in 
$C$-representation \cite{DD95} are given 
in Table \ref{param}. They vary very much with 
different Skyrme forces, e.g., the coupling 
constant $C^s_0(\rho=0)$ is 14, 84 and --208 for 
the respective SIII, SkI3 and SLy4 forces.

The polarization energy 
depends strongly on the blocked level by 
an odd neutron. Even the sign of energy 
changes with different choice of the 
blocked level. In the present study 
we simply choose as the blocked level 
the lowest quasi-particle state that 
is obtained from the HFBE calculation.

\begin{figure}[tpb]
\epsfxsize=.8\textwidth
\centerline{\epsffile{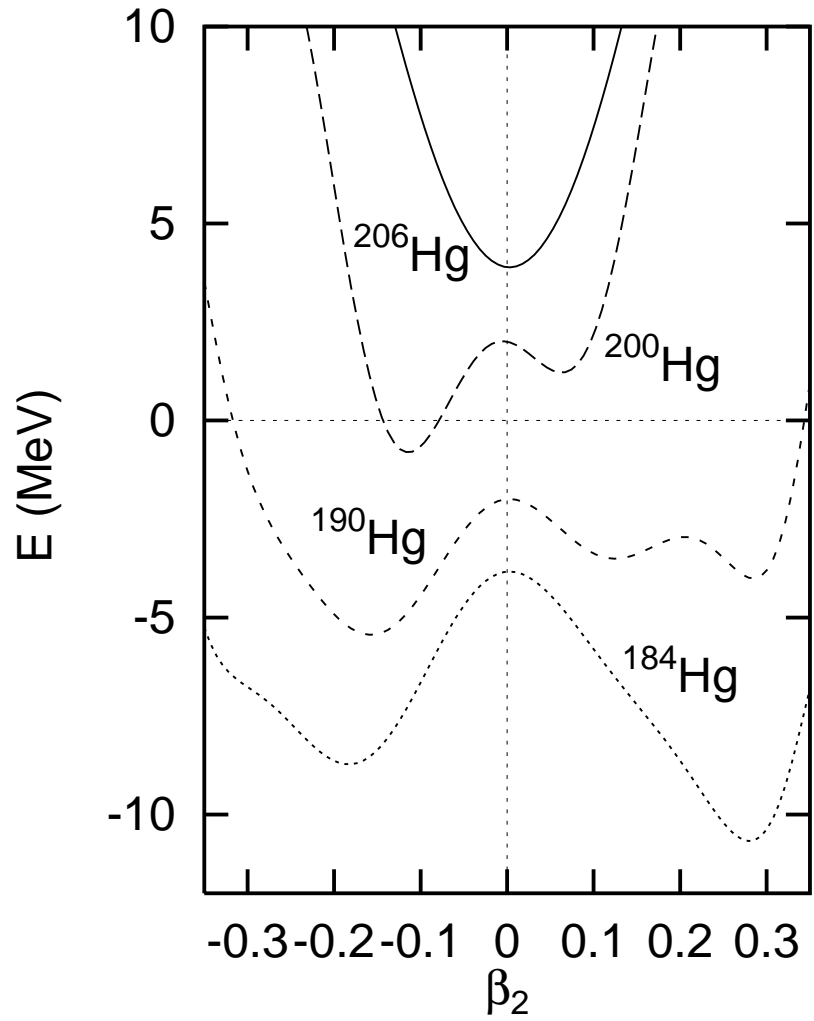}}
\caption{\label{PES}
Deformation energy of Hg isotopes calculated by using 
the constraint SIII-HF method.
}
\end{figure}

Mercury isotopes are very interesting 
because the shape of the ground state 
varies very much as the neutron number 
decreases from the closed shell. 
Figure \ref{PES} shows the deformation 
energy of even Hg isotopes calculated 
by the constraint SIII-HF method. 
We see the spherical shape in the 
neutron-closed $^{206}$Hg. 
The oblate minimum with deformation $\beta_2\sim-0.17$ 
then develops in $^{200}$Hg. 
A rather complicated prolate minimum appears in $^{190}$Hg. 
Finally the deep prolate minimum 
with deformation $\beta_2\sim0.29$ 
develops in $^{184}$Hg. 
The oblate-prolate shape coexistence 
in light Hg isotopes has been predicted 
by many authors, e.g., \cite{FP75,CL73,YT97,RR00,NV02,FG72,BB87}.

\begin{figure}[htpb]
\epsfxsize=.8\textwidth
\centerline{\epsffile{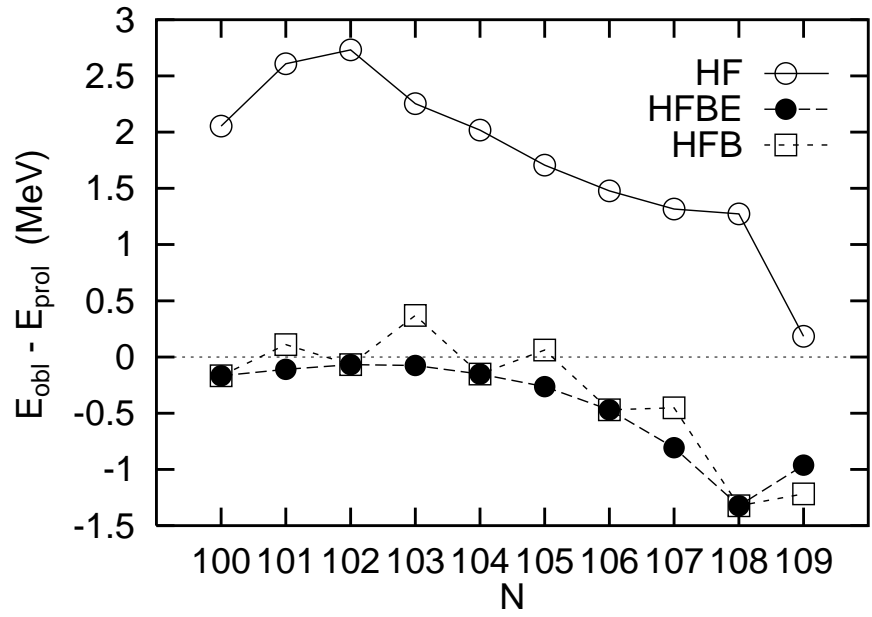}}
\caption{\label{p-o}
The oblate-prolate energy difference as a function of 
the neutron number $N$ of Hg isotopes.
}
\end{figure}

Figure \ref{p-o} shows the 
SIII calculation of the oblate-prolate energy difference. 
The HF calculation (white circle) predicts 
prolate deformation in the ground state of all Hg isotopes. 
The HFBE calculation (black circle), 
the fully paired state without the blocking effect, 
predicts oblate deformation in the ground state 
of all Hg isotopes. 
The shape changes due to 
the large gain of pairing energy in 
the oblate minimum because of the high level density 
as compared to the level density 
in the prolate minimum. 
In the SIII-HFB calculation (white square), 
on the other hand, 
the shape of odd mass $^{181-185}$Hg isotopes 
returns to the prolate shape. 
This is because the loss of pairing energy due to 
the blocking by the odd neutron is also large 
in the oblate minimum than in the prolate one. 
The effect of the time-odd mean fields on 
the oblate-prolate transition is weak 
and a few tenths of the blocking effect.

\begin{figure}[htbp]
\epsfxsize=.8\textwidth
\centerline{\epsffile{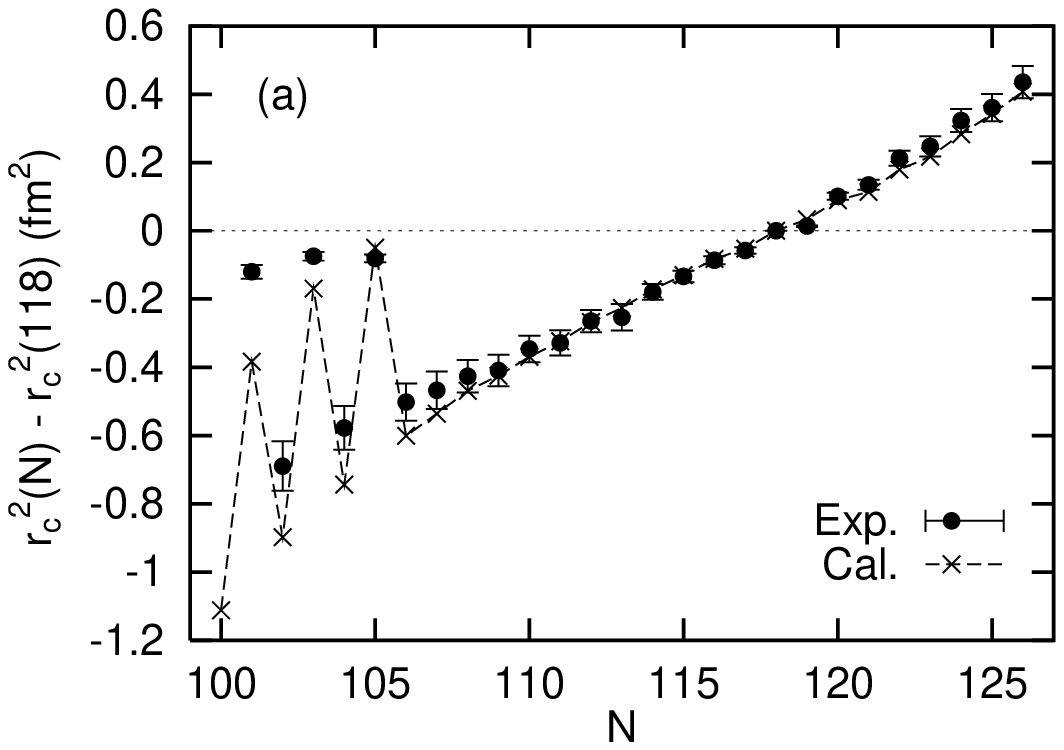}}
\epsfxsize=.8\textwidth
\centerline{\epsffile{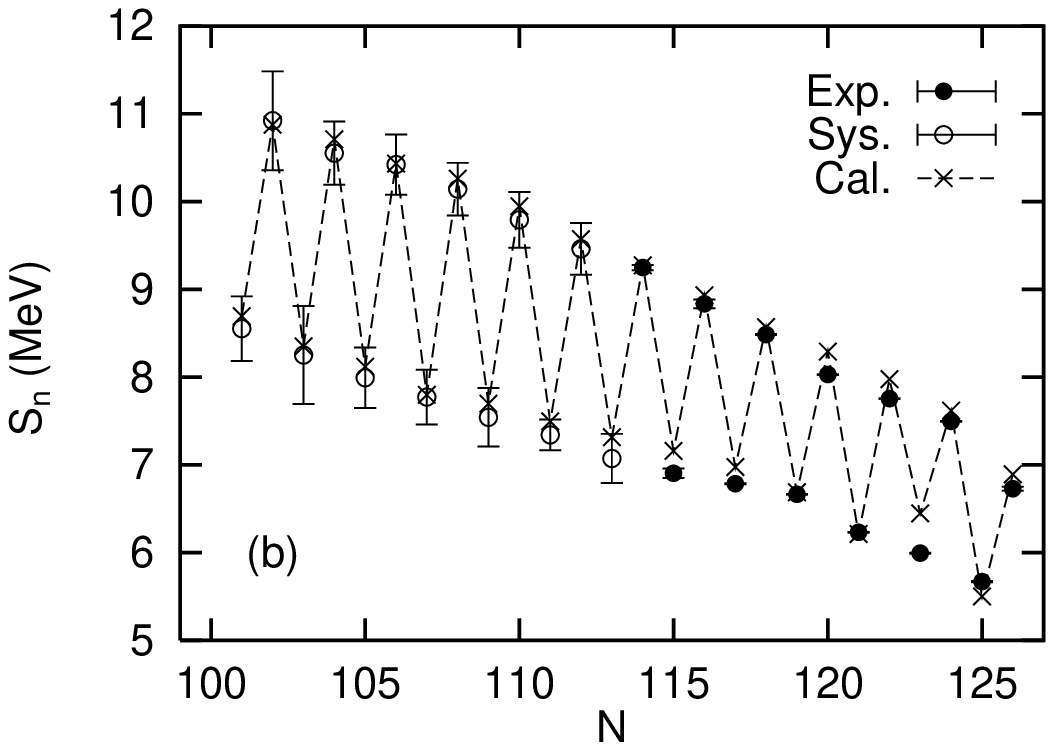}}
\caption{\label{IS}
(a) The mean-square charge radius of $^{180-206}$Hg 
with respect to $^{198}$Hg. 
(b) The one-neutron separation energy as a function of 
the neutron number $N$ of Hg isotopes.
Experiments are taken from {\protect \cite{AH87}}
and {\protect \cite{AW95}}, respectively.
White circles in (b) denote the separation energy derived 
from systematic trends 
{\protect \cite{AW95}}. 
}
\end{figure}

Figure \ref{IS} shows the SIII-HFB calculation of 
(a) the mean-square charge radius and 
(b) one-neutron separation energy of $^{180-206}$Hg isotopes. 
The sharp increase in the mean-square 
radius of the odd-mass $^{181-185}$Hg isotopes 
is very well explained by the calculation. 
The one-neutron separation energy is also 
well reproduced in the calculation.
This is so because the depth of the pairing force 
has been adjusted to reproduce the observed separation energies. 
It is noted that the white circle in Fig.~\ref{IS}(b) 
is a derived value from systematic trends \cite{AW95} 
and not an experiment.

As is well known, a proper description of odd nuclei 
by mean field theory requires to break 
the time-reversal symmetry and introduces 
the time-odd components (TOC) in the mean field. 
The effect of an odd neutron on the HFB energy is 
therefore twofold; (a) The loss of pairing energy 
due to the blocking of the Fermi level by the odd neutron, 
and (b) the time-odd mean-field energy arising from 
the blocked level. The time-odd mean fields 
have been studied both in the analysis of 
super-deformed rotational bands \cite{DD95} 
and the odd-even nuclear mass difference \cite{DB01}.

In order to see the blocking energy 
(the loss of pairing energy) 
and the polarization energy 
(the time-odd mean-field energy) separately, 
we define the blocking energy $E^{\rm block}$ 
as the energy difference between $E^{\rm HFB}$ 
with no TOC and $E^{\rm HFBE}$. 
We also define the polarization energy 
$E^{\rm pol}$ as the energy difference 
between $E^{\rm HFB}$ and $E^{\rm HFB}$ with no TOC, i.e., 
\begin{eqnarray}
E^{\rm block} &=&
 E^{\rm HFB}(\mbox{\rm no TOC}) - E^{\rm HFBE} , \\
E^{\rm pol}   &=& E^{\rm HFB} - E^{\rm HFB}(\mbox{\rm no TOC}) .
\end{eqnarray}

\begin{figure}[htbp]
\begin{center}
\begin{minipage}{.45\textwidth}
\epsfxsize=\textwidth
\centerline{\epsffile{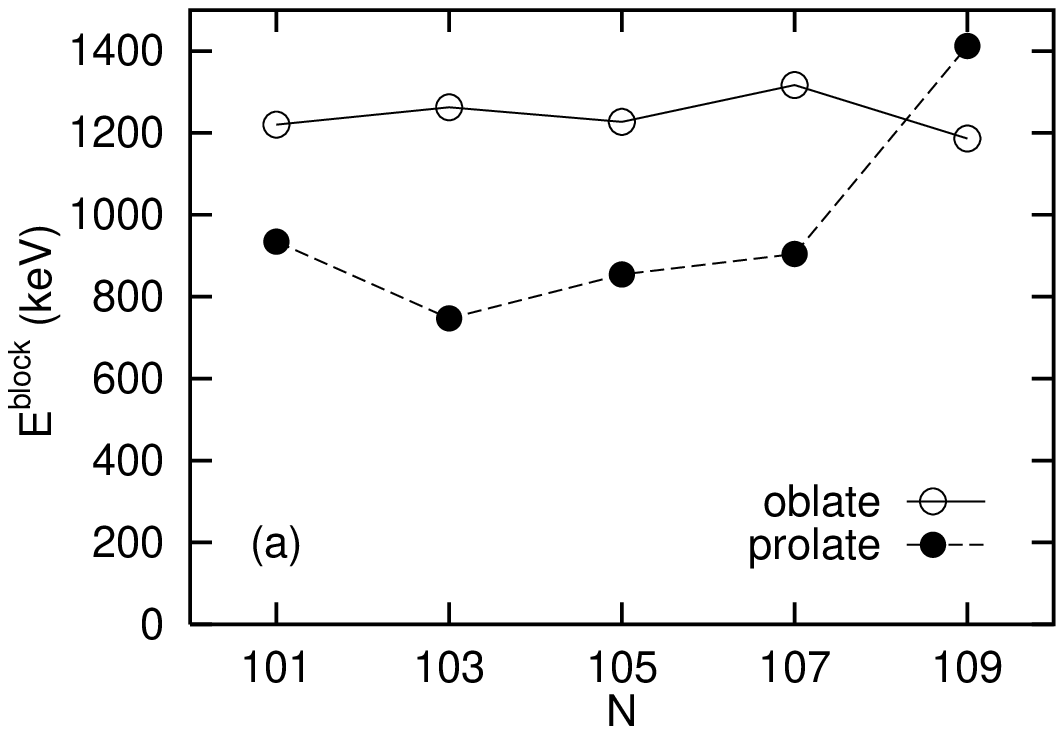}}
\epsfxsize=\textwidth
\centerline{\epsffile{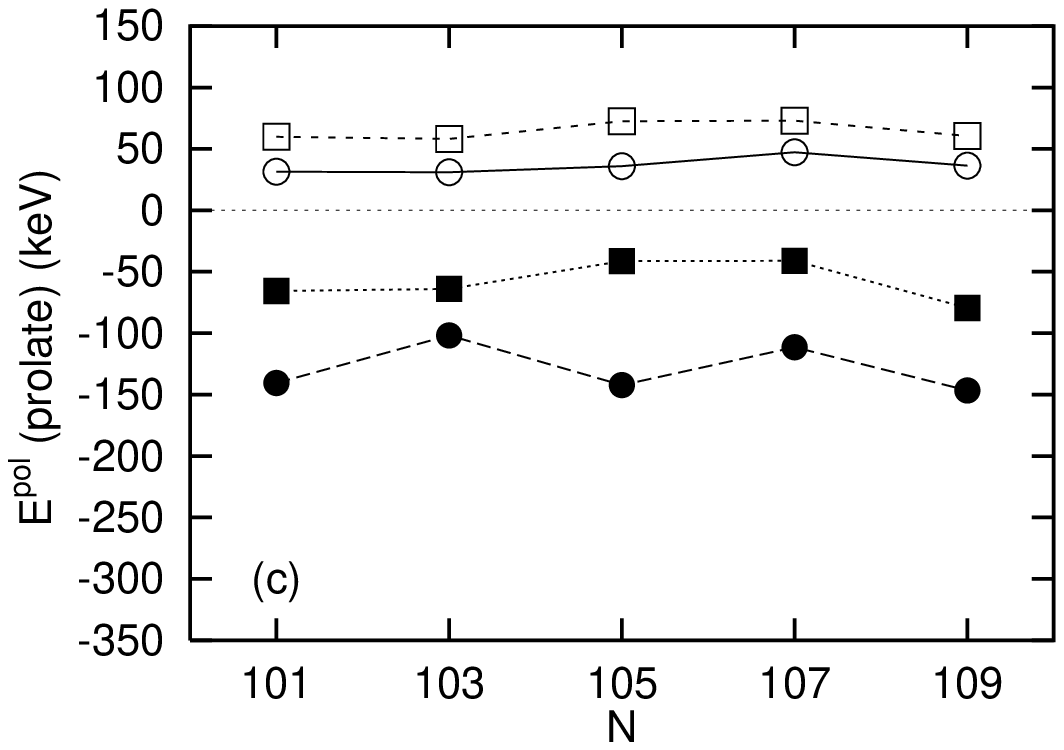}}
\end{minipage}
\begin{minipage}{.45\textwidth}
\epsfxsize=\textwidth
\centerline{\epsffile{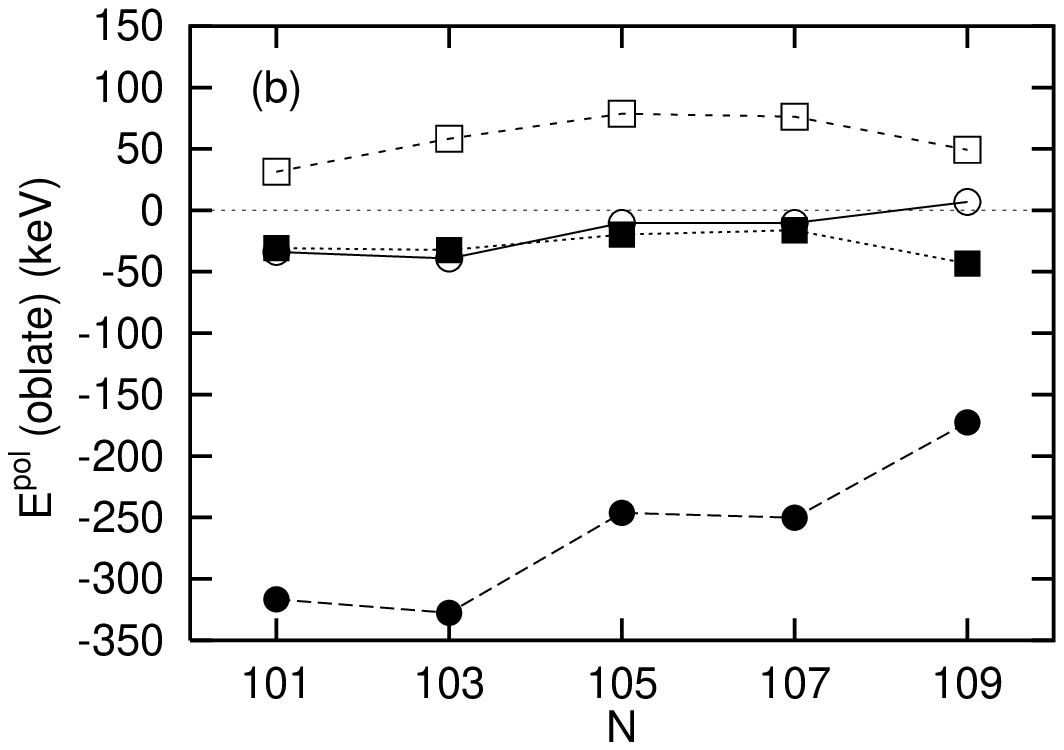}}
\epsfxsize=\textwidth
\centerline{\epsffile{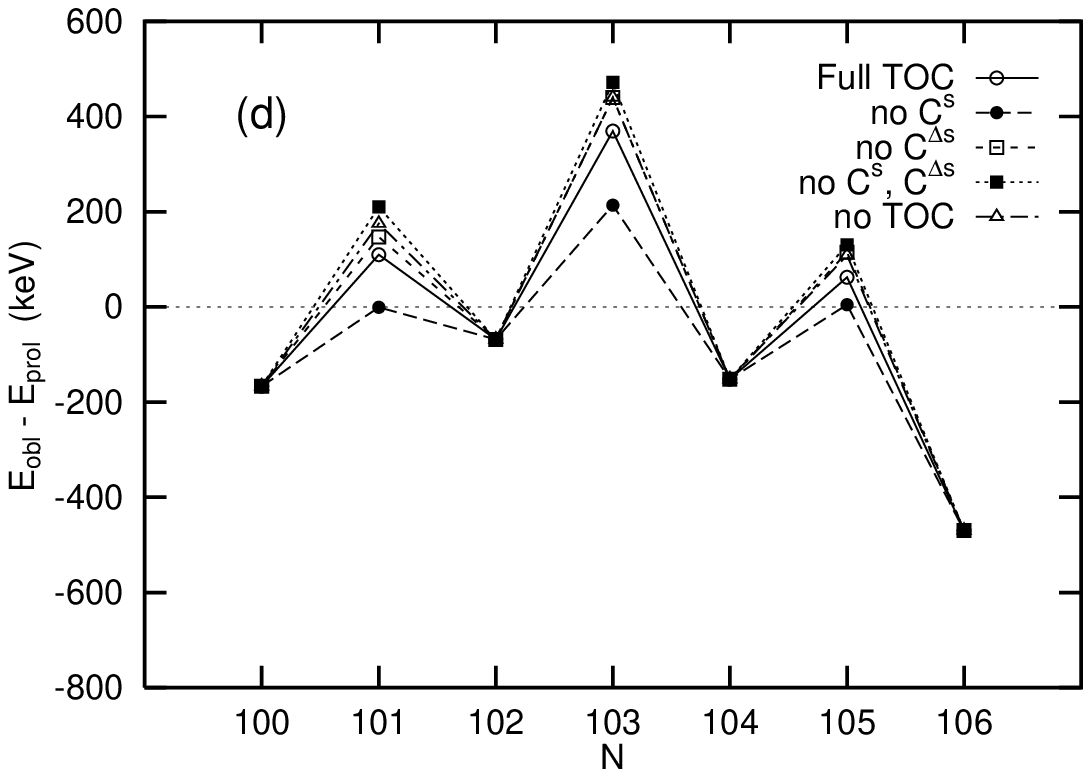}}
\end{minipage}
\end{center}
\caption{\label{p-o_SIII}
Results of the SIII-HFB calculation
with the pairing parameters of $V_0=-567$ MeV and 
$\rho_{\rm c}=0.145$ fm$^{-3}$. 
(a) The loss of pairing energy due to 
the self-consistent blocking of the Fermi level by the odd neutron. 
(b) The polarization energy in the oblate minimum. 
(c) The polarization energy in the prolate minimum. 
(d) The oblate-prolate energy difference. 
Case (i): the calculation with full time-odd terms (white circle). 
Case (ii): the calculation without the $C^s_t$ terms (black circle). 
Case (iii): the calculation without the $C^{\Delta s}_t$ terms (white square). 
Case (iv): the calculation without
           both  $C^s_t$ and $C^{\Delta s}_t$ terms (black square). 
The calculation with no time-odd components is shown in white triangle.  
}
\end{figure}

Figure \ref{p-o_SIII} shows the SIII-HFB calculation of 
(a) the blocking energy in the oblate and prolate minimum, 
(b) the polarization energy in the oblate minimum, 
(c) the polarization energy in the prolate minimum and 
(d) the oblate-prolate difference of HFB energy. 
In Fig.~\ref{p-o_SIII}(a), the blocking energy 
in the oblate minimum is larger by $\sim$200 keV 
than the one in the prolate minimum. 
This results from higher neutron level-density 
in the former than in the latter. 
The blocking energies are reversed in $^{189}$Hg 
because of the small deformation $\beta=0.08$ 
in the prolate minimum of this isotope and 
the level density in the minimum is high accordingly.

As mentioned in Section 2.2,
the $\Cs{t}$ and $\Cds{t}$ terms are
free from the time-even parameterization.
To investigate their effects on the HFB energy,
we have further divided the polarization energy into 
contributions with (i) the full time-odd terms, 
(ii) the time-odd terms without $C^s_t$, 
(iii) the time-odd terms without $C^{\Delta s}_t$, 
and (iv) the time-odd terms without both $C^s_t$
 and $C^{\Delta s}_t$. 
In the following we assume simultaneous 
modifications of the isoscalar 
and isovector coupling constants of a given species.

Figures~\ref{p-o_SIII} (b) and (c) display 
the contributions to the polarization energy. 
The polarization energy with full time-odd terms 
(white circle) is quite small. 
It is attractive and less than --50 keV 
in the oblate minimum, while it is repulsive 
and less than 50 keV in the prolate minimum. 
When the $C^s_t$ terms are omitted (black circle), 
the polarization energy becomes attractive 
and more than --250 keV in the oblate minimum 
and it is attractive around --150 keV in the prolate minimum. 
The $C^s_t$ terms have repulsive contribution to the HFB energy. 
When the $C^{\Delta s}_t$ terms are omitted (white square), 
on the other hand,
the polarization energy becomes repulsive and $\sim$50 keV 
in both minima. 
The $C^{\Delta s}_t$ terms have large attractive 
contributions to the HFB energy. 
When both $C^s_t$ and $C^{\Delta s}_t$ terms 
are omitted (black square), 
the polarization energy becomes small. 
The effect of the $C^s_t$ and $C^{\Delta s}_t$ 
terms is opposite and 
cancels with each other.

Figure \ref{p-o_SIII}(d) shows the difference of 
energy in the oblate and prolate minimum. 
The figure shows the calculation 
with the full time-odd terms (white circle), 
without the $C^s_t$ terms (black circle), 
without the $C^{\Delta s}_t$ terms (white square), 
without both $C^s_t$ and $C^{\Delta s}_t$ terms (black square) 
and without all time-odd terms (white triangle). 
When both $C^s_t$ and $C^{\Delta s}_t$ terms are omitted, 
the oblate shape is predicted in $^{181}$Hg, 
while the prolate shape is calculated in $^{185}$Hg 
with marginal energy difference between 
the oblate and prolate minima.

Table \ref{p-o_SIIIt} summarizes 
the oblate-prolate difference of the HFBE energy
$\Delta E^{\rm HFBE}$, 
the blocking energy $\Delta E^{\rm block}$, 
the polarization energy $\Delta E^{\rm pol}$ 
and the sum of these contributions $\Delta E^{\rm HFB}$. 
The polarization energy $\Delta E^{\rm pol}$ is 
calculated with four different time-odd mean fields. 
They are the HFB calculation 
(i) with full time-odd terms, 
(ii) without the $C^s_t$ terms,
(iii) without the $C^{\Delta s}_t$ terms and 
(iv) without both $C^s_t$ and $C^{\Delta s}_t$ terms.

The energy difference $\Delta E^{\rm HFBE}$ shows 
that the minimum is found at the neutron number $N=102$ 
and the oblate minimum gets deeper than 
the prolate one as the neutron number is 
away from $N=102$ in both directions. 
For the blocking energy $E^{\rm block}$, 
it is repulsive and larger by 400 keV 
in the oblate minimum than in the prolate one. 
The blocked level is labeled with the 
quantum number $\Omega$, the component of 
neutron angular momentum along the symmetry axis.

For the odd mass $^{181-185}$Hg, the blocked 
level has $\Omega=1/2$ in the oblate minimum, 
while it has $\Omega=5/2$ or $7/2$ in the prolate minimum. 
In the calculation with full time-odd terms (case (i)), 
the polarization energy is attractive in the oblate minimum, 
while it is repulsive in the prolate minimum, 
leading to the the attractive energy difference 
of $\Delta E^{\rm pol}$. 
The attractive $\Delta E^{\rm pol}$ weakens 
the repulsive $\Delta E^{\rm block}$ 
and prevents the oblate-prolate shape transition 
in these isotopes. 
In the calculation without the $C^s_t$ terms (case (ii)),
the polarization energy becomes attractive 
because of the absence of the repulsive $C^s_t$ terms. 
In this case, the polarization energy $\Delta E^{\rm pol}$ 
almost cancels the blocking energy $\Delta E^{\rm block}$. 
In the calculation without the
$C^{\rm \Delta s}_t$ terms (case (iii)), 
the polarization energy is weakly repulsive because of 
the absence of strongly attractive $C^{\Delta s}_t$ terms. 
Finally in the calculation 
without both $C^s_t$ and $C^{\Delta s}_t$ terms, 
the polarization energy is weakly attractive in both minima. 
The effects of the $C^s_t$ and $C^{\rm \Delta s}_t$ terms on 
the polarization energy will be discussed more generally
in the last paragraph of this section.

\begin{figure}[htbp]
\begin{center}
\begin{minipage}{.45\textwidth}
\epsfxsize=\textwidth
\centerline{\epsffile{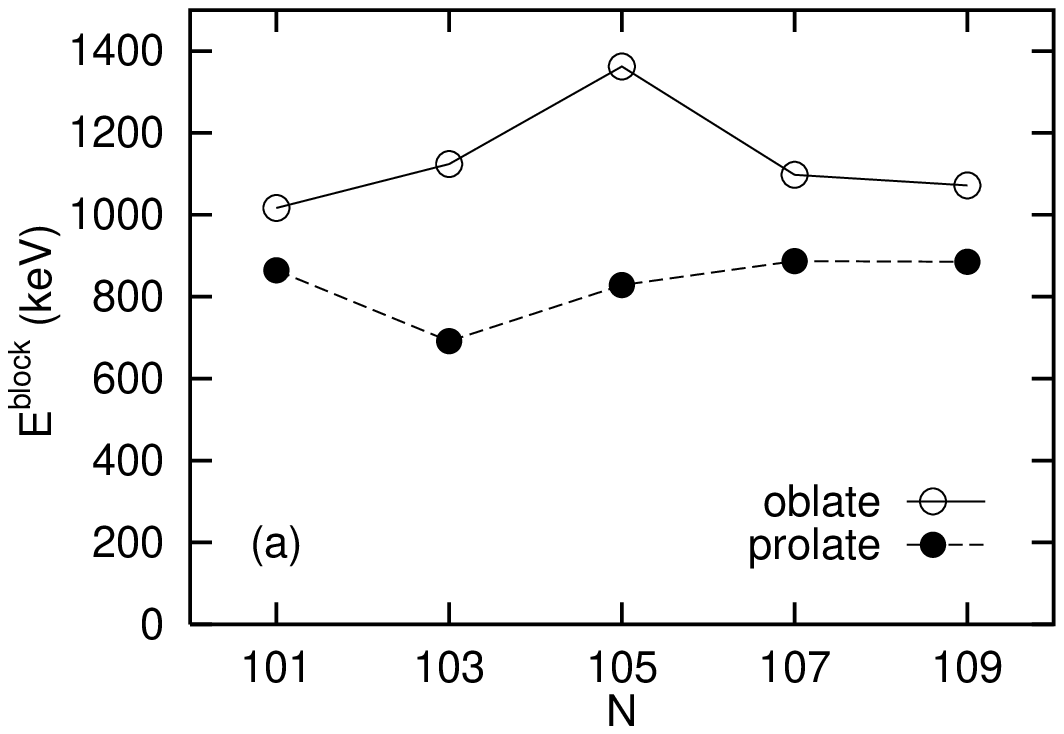}}
\epsfxsize=\textwidth
\centerline{\epsffile{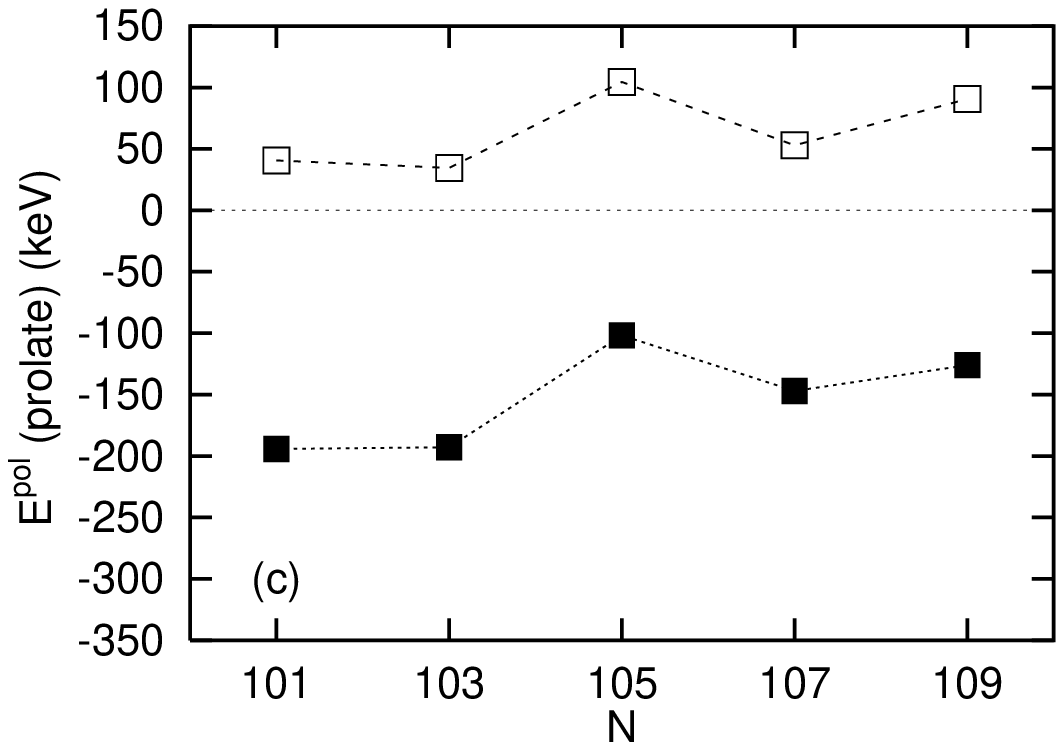}}
\end{minipage}
\begin{minipage}{.45\textwidth}
\epsfxsize=\textwidth
\centerline{\epsffile{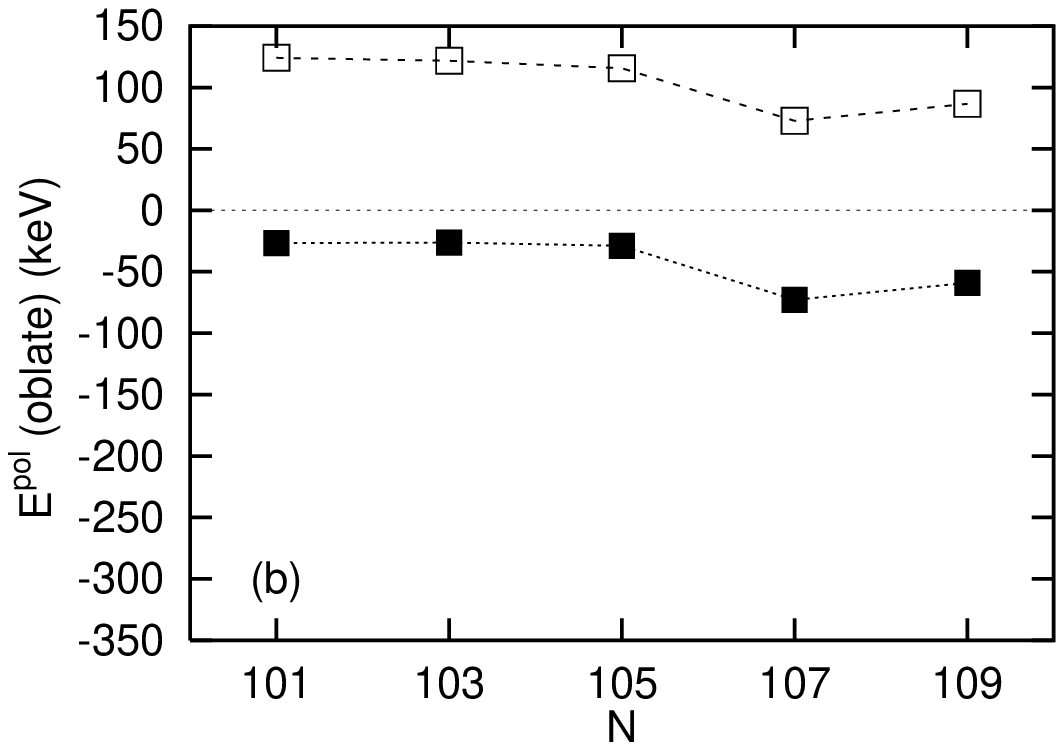}}
\epsfxsize=\textwidth
\centerline{\epsffile{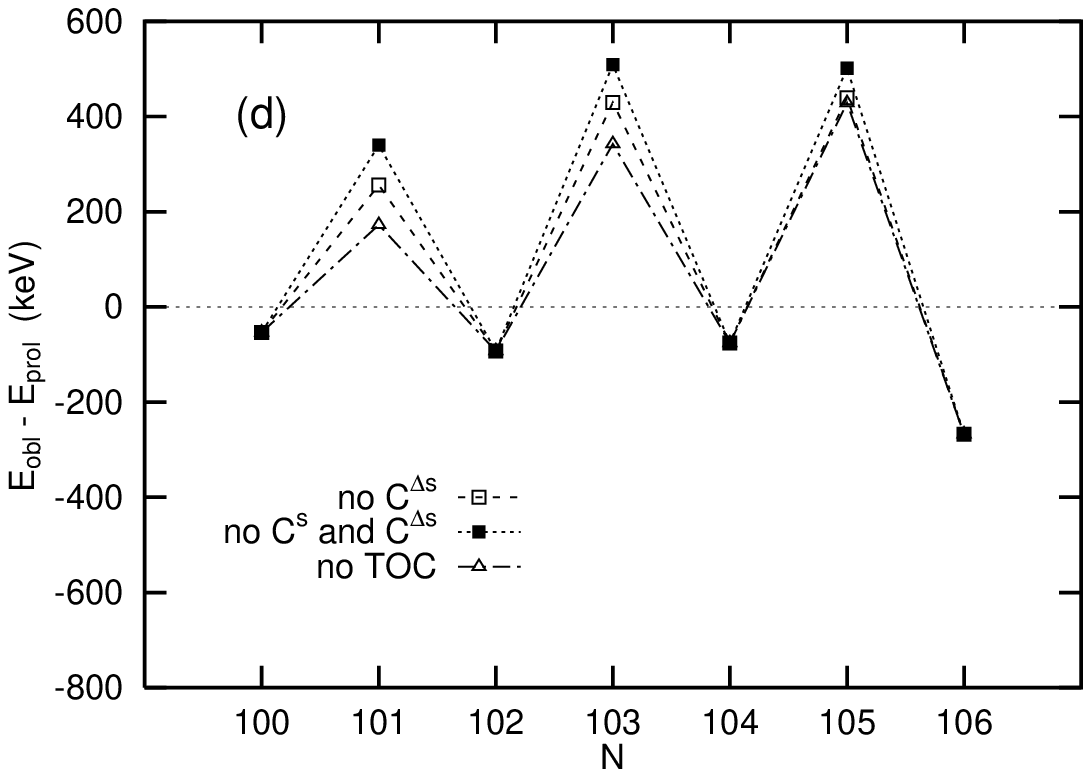}}
\end{minipage}
\end{center}
\caption{\label{p-o_SkI3}
Results of the SkI3-HFB calculation
with the pairing parameters of $V_0=-560$ MeV 
and $\rho_{\rm c}=0.158$ fm$^{-3}$.  
(a) The loss of pairing energy due to 
the self-consistent blocking of the Fermi level by the odd neutron. 
(b) The polarization energy in the oblate minimum. 
(c) The polarization energy in the prolate minimum. 
(d) The oblate-prolate energy difference. 
Case (iii): the calculation without the $C^{\Delta s}_t$ terms (white square). 
Case (iv): the calculation without both 
           $C^s_t$ and $C^{\Delta s}_t$ terms (black square). 
The calculation with no time-odd components is also shown 
in white triangle.
}
\end{figure}

Figure~\ref{p-o_SkI3} shows the results of 
the SkI3-HFB calculation. 
We had the difficulty of convergence 
in this calculation when the full time-odd components are considered.
The difficulty arises mainly due to 
the strong coupling-constant $C^{\Delta s}_0$ derived from 
the SkI3 parameters (see Table \ref{param}) and 
partly due to the large HF basis
of the present study ($N_0=18$). 
The difficulty occurs when the energy 
splitting between the time-reversed partners becomes large
and a few of the quasi-particle states have negative energy.
Since we have no way to correct this difficulty,
we simply omit the $C^{\Delta s}_t$ terms
and display the results in Fig.~\ref{p-o_SkI3}. 
The blocking energy $E^{\rm block}$ in Fig.~\ref{p-o_SkI3}(a)
is similar to the one in the SIII-HFB calculation. 
Figures~\ref{p-o_SkI3}(b) and \ref{p-o_SkI3}(c) show 
the polarization energy (iii) 
without $C^{\Delta s}_t$ terms (white square) 
and (iv) without both $C^{\Delta s}_t$ and $C^s_t$ terms (black square). 
The $C^s_t$ terms have repulsive contributions of $\sim$150 keV 
in the oblate minimum while it is $\sim$200 keV 
in the prolate minimum. In Fig.~\ref{p-o_SkI3}(d), 
the oblate-prolate transition is very well 
explained when the terms $C^{\Delta s}_t$ are omitted. 
It is also well explained by omitting the whole time-odd components.

Table \ref{p-o_SkI3t} summarizes the results of 
the SkI3-HFB calculation. 
The energy difference $\Delta E^{\rm HFBE}$ is 
nearly the same for even and odd $^{180-185}$Hg isotopes.
This fact helps very much in reproducing 
the oblate-prolate transition in 
the odd mass $^{181-185}$Hg isotopes. 
The SkI3 force is very well suited to describe
the mean fields of light Hg isotopes.

\begin{figure}[htbp]
\begin{center}
\begin{minipage}{.45\textwidth}
\epsfxsize=\textwidth
\centerline{\epsffile{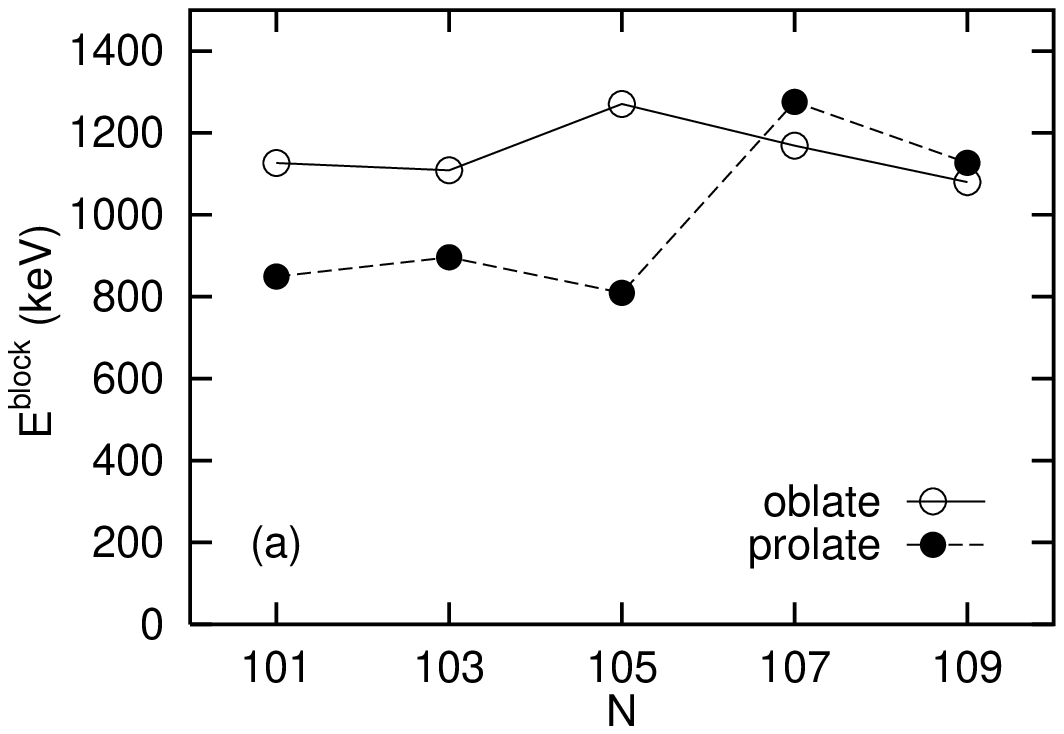}}
\epsfxsize=\textwidth
\centerline{\epsffile{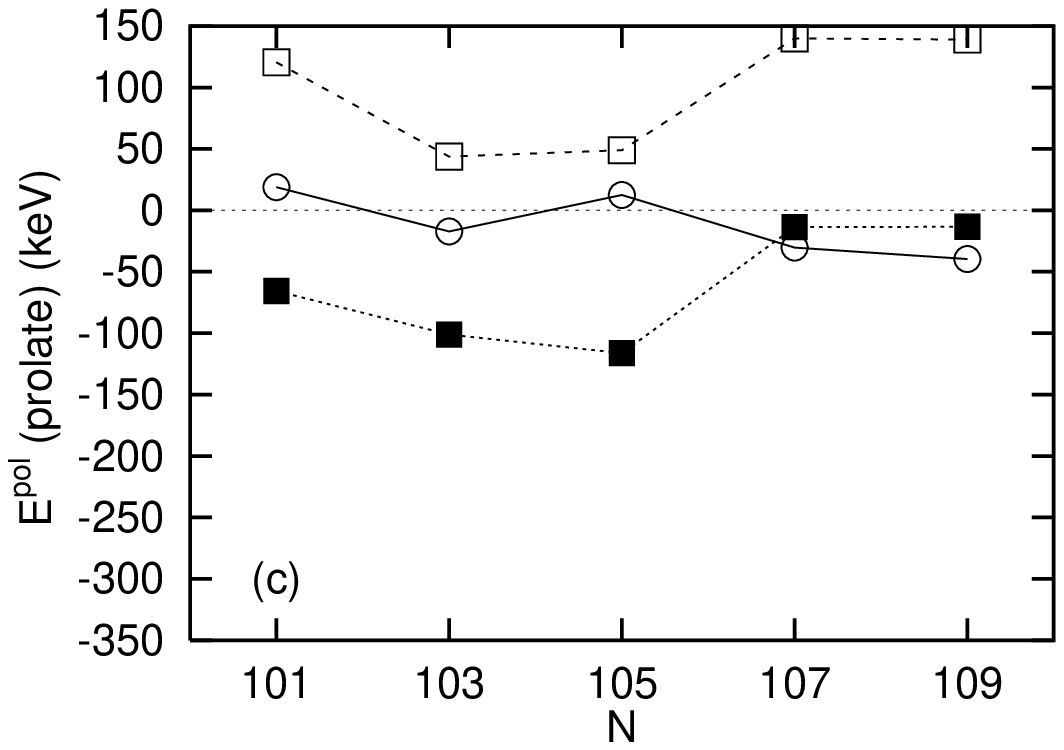}}
\end{minipage}
\begin{minipage}{.45\textwidth}
\epsfxsize=\textwidth
\centerline{\epsffile{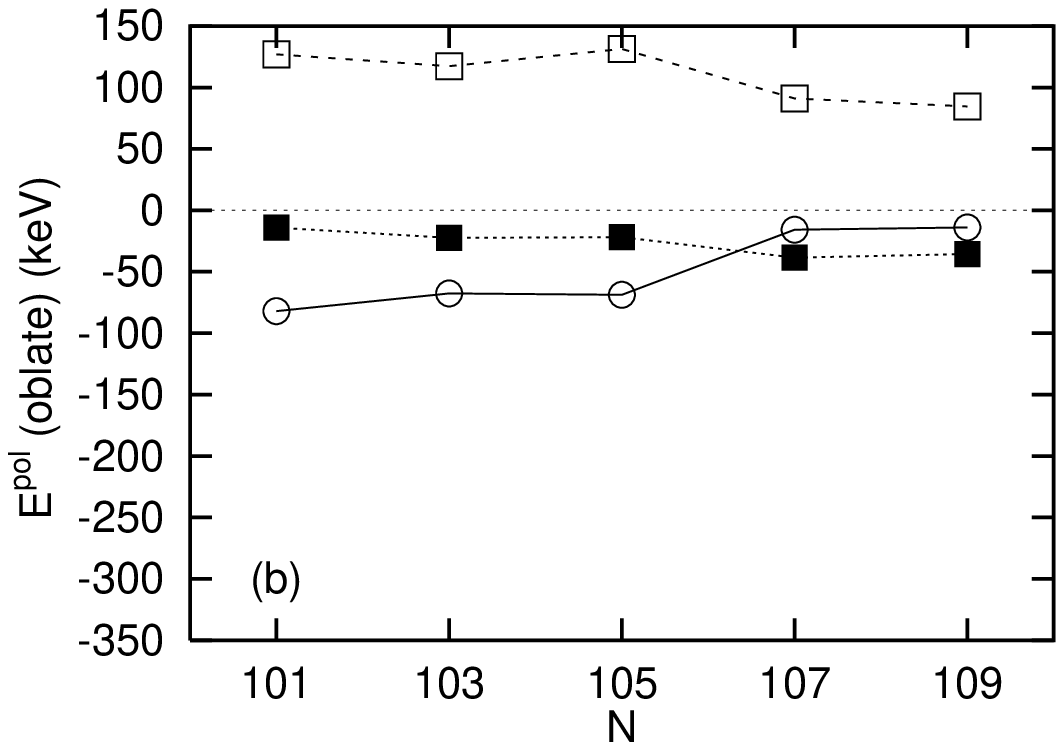}}
\epsfxsize=\textwidth
\centerline{\epsffile{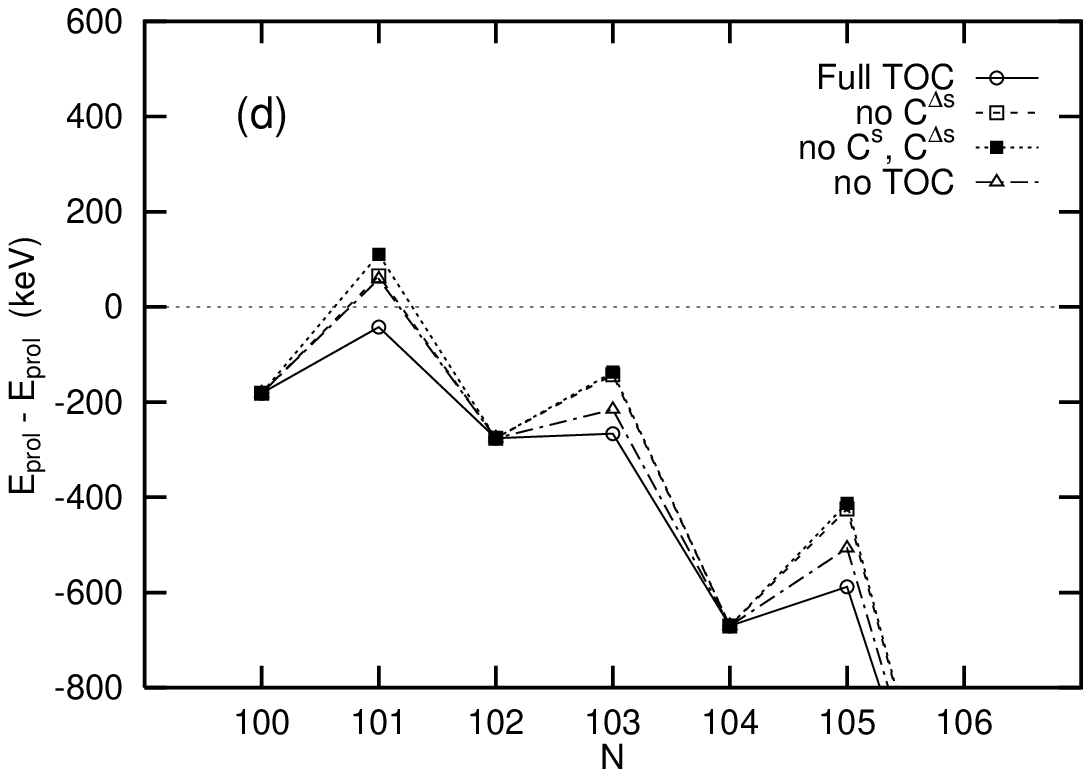}}
\end{minipage}
\end{center}
\caption{\label{p-o_SLy4}
Results of the SLy4-HFB calculation
with the pairing parameters of $V_0=-500$ MeV 
and $\rho_{\rm c}=0.160$ fm$^{-3}$.
(a) The loss of pairing energy due to 
the self-consistent blocking of the Fermi level by the odd neutron. 
(b) The polarization energy in the oblate minimum. 
(c) The polarization energy in the prolate minimum. 
(d) The oblate-prolate energy difference. 
Case (i): the calculation with full time-odd terms (white circle). 
Case (iii): the calculation without the $C^{\Delta s}_t$ terms (white square). 
Case (iv): the calculation without both 
           $C^s_t$ and $C^{\Delta s}_t$  terms (black square). 
The calculation with no time-odd components is also shown 
in white triangle. 
}
\end{figure}

Figure~\ref{p-o_SLy4} shows the results of the SLy4-HFB calculation. 
The blocking energy $E^{\rm block}$ in the odd-mass $^{181-185}$Hg 
is larger by $\sim$200 keV in the oblate minimum 
than in the prolate one. 
However, their magnitudes are reversed 
at the neutron number $N=107$ and $N=109$. 
The reason is the same as the SIII calculation. 
The equilibrium deformation in the prolate minimum 
is reduced to $\beta = 0.12$ in these isotopes. 
Figures~\ref{p-o_SLy4}(b) and \ref{p-o_SLy4}(c) 
show the terms $C^s_t$ have repulsive contributions 
to the polarization energy. 
In addition, the terms $C^s_t$ and $C^{\Delta s}_t$ 
have opposite signs and nearly cancel with each other.

The oblate-prolate energy difference 
in Fig.~\ref{p-o_SLy4}(d) shows that 
the SLy4 force is not very well suited
to describe deformation properties of light Hg isotopes. 
Here we also had the difficulty of convergence
when the $C^s_t$ terms are not included in the calculation.
The difficulty may be due to the $C^{\Delta s}_t$ terms.  
The repulsive $C^s_t$ terms cancel 
the attractive $C^{\Delta s}_t$ terms. 
Without the $C^s_t$ terms, 
the attractive $C^{\Delta s}_t$ terms are quite strong 
and cause the difficulty of convergence.

Table \ref{p-o_SLy4t} summarizes the results of the SLy4-HFB calculation. 
The energy difference $\Delta E^{\rm HFBE}$ shows that 
the SLy4 force favors the oblate shape 
in light Hg isotopes much more than the SIII and SkI3 forces. 
This fact results in the oblate-prolate transition 
only in the ground state of $^{181}$Hg isotope. 
The oblate shape is predicted for all heavier isotopes.

\begin{figure}[htbp]
\begin{center}
\epsfxsize=.45\textwidth
\centerline{\epsffile{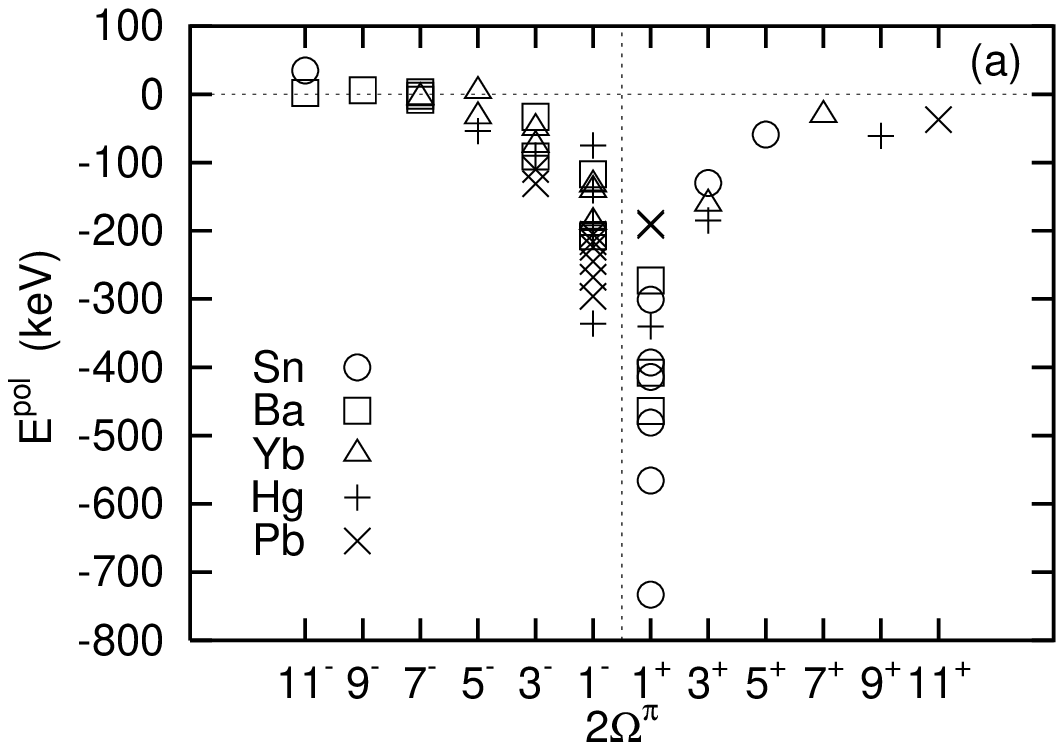}}
\epsfxsize=.45\textwidth
\centerline{\epsffile{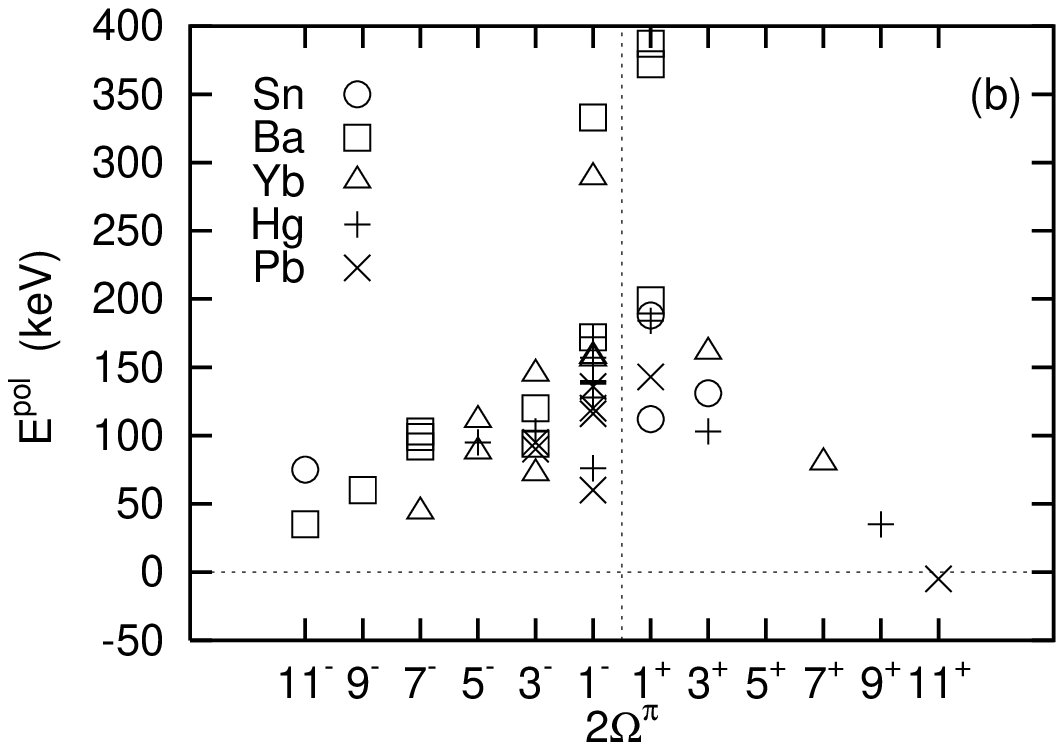}}
\epsfxsize=.45\textwidth
\centerline{\epsffile{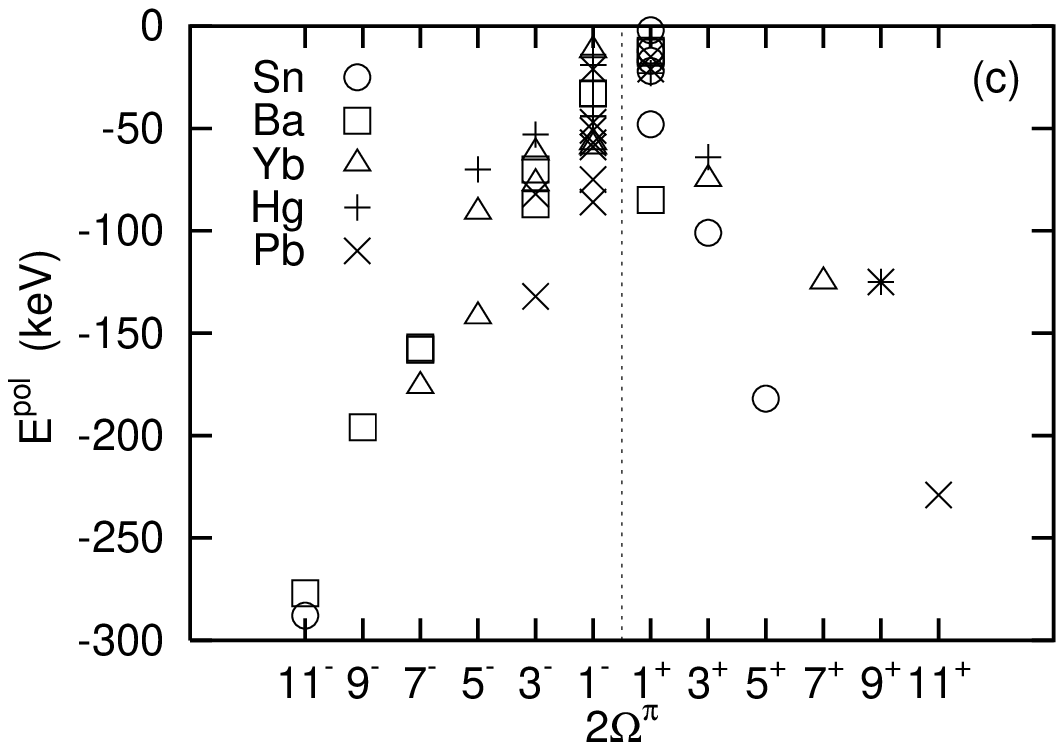}}
\end{center}
\caption{\label{test}
The time-odd mean-field energy calculated by
the SkI3-HFB method with the basis space of $N_0=10$.
The energy is plotted against $2\Omega^\pm$, 
twice of the component of neutron angular momentum
along the symmetry axis.
(a) The calculation with full time-odd terms, 
(b) the calculation without the $C^{\Delta s}_t$ terms, and 
(c) the calculation without both $C^{\Delta s}_t$ and $C^s_t$ terms. 
}
\end{figure}

To see the effects of the time-odd 
$C^s_t$ and $C^{\Delta s}_t$ terms more closely, 
we have analysed the odd-mass Sn, Ba, Yb, Hg 
and Pb isotopes by the SkI3-HFB method 
with the basis space of $N_0=10$. 
A small basis space was chosen in order 
to avoid the difficulty of convergence. 
Results of the SkI3-HF calculation are shown in Fig.~\ref{test}. 
The polarization energy is plotted against $2\times\Omega^\pi$ 
of the blocked level. 
Polarization energies are calculated
(a) with the full time-odd terms, 
(b) without $C^{\Delta s}_t$ terms and 
(c) without both $C^{\Delta s}_t$ and $C^s_t$ terms. 
We may conclude that the $C^{\Delta s}_t$ terms are 
attractive and have a large effect on the blocked 
level with low $\Omega$, in particular $\Omega^\pi=1/2^\pm$. 
The terms $C^s_t$ are repulsive for levels with any 
value of $\Omega$. 
The remaining time-odd terms are 
attractive for levels with high $\Omega$ and are 
negligible for levels with low $\Omega$. 
The same features of the time-odd terms are
observed also for the SIII and SLy4 interactions.
For the odd-mass $^{181-185}$Hg 
in Tables \ref{p-o_SIIIt}, \ref{p-o_SkI3t} and \ref{p-o_SLy4t},
we have seen that
the time-odd terms have attractive contributions to
$\Delta E^{\rm HFB}$ in the cases (i) and (ii),
while the terms have repulsive contributions in the case (iv).

\section{Conclusions}

The shape and the mean-square radius of Hg isotopes 
have been investigated by using 
the Skyrme Hartree-Fock-Bogoliubov method. 
We have first examined effective 
interactions SIII, SGII, SkM$^*$, SKX, SkI1, SkI2, 
SkI3, SkI4, SkI5 and SLy4 to see
if they can predict the oblate-prolate shape 
coexistence in light Hg isotopes. 
We have then examined 
the loss of pairing energy due to the odd neutron 
by employing the zero-range density-dependent pairing force. 
Lipkin-Nogami corrections to both even and odd nuclei are 
calculated within the HFB formalism. 
We have also investigated the effect of the time-odd $C^{\Delta s}_t$ 
and $C^s_t$ terms on the HFB energy in relation to 
the blocked level by the odd neutron.

We have obtained the following conclusions. 
(1) The effective interactions 
SIII and SkI3 have desirable deformation properties 
in predicting the oblate-prolate shape coexistence 
in light Hg isotopes. 
(2) The blocking energy of the pairing 
is the most important factor to change 
the shape of the odd-mass $^{181-185}$Hg isotopes. 
(3) The effect of the time-odd terms on 
the prolate-oblate transition is weak 
and cancels the effect of the blocking energy by a few tenths.

We have also investigated the effects of 
the time-odd terms $C^s_t$ and $C^{\Delta s}_t$ 
by blocking several levels near the Fermi surface of 
Sn, Ba, Yb, Hg and Pb isotopes. 
We have found that the terms $C^{\Delta s}_t$ are 
attractive for the blocked levels with low $\Omega$, 
in particular $\Omega^\pi=1/2^\pm$. 
On the other hand, the terms $C^s_t$ are 
repulsive for the blocked levels with all values of $\Omega$. 
The remaining time-odd terms are attractive for 
the blocked levels with high $\Omega$ and negligible 
for the blocked levels with low $\Omega$.

We had the difficulty of convergence in 
the SkI3-HFB calculation 
when the $C^{\Delta s}_t$ terms are included. 
The difficulty also arises from the 
SLy4-HFB calculation when the $C^s_t$ terms are omitted. 
These difficulties are mainly due to 
the strong $C^{\Delta s}_t$ terms 
and partly due to the large basis space of 
the calculation ($N_0=18$). 
For large $C^{\Delta s}_t$, 
the energy splitting 
between the time-reversed partners becomes large
and a few of the quasi-particle states have negative energy.

\section*{Acknowledgments}

We would like to thank K. Arita for many illuminating discussions. 
Numerical computation in this work was partly carried out 
at the Yukawa-Institute Computer Facility, Kyoto University.


\clearpage
\begin{table}[htbp]
\caption{\label{param}
The time-odd coupling constants $C^s_t$ and $C^{\Delta s}_t$ for
the Skyrme SIII, SkI3 and SLy4 forces.
The coupling constants $C^s_t$ are given in MeV$\cdot$fm$^3$, while
$C^{\Delta s}_t$ are given in MeV$\cdot$fm$^5$.
}
\begin{center}
\begin{tabular}{lcccccc}
\hline
Force  & $C_0^{s}(\rho=0)$
       & $C_0^{s}(\rho=\rho_{\rm NM})$
       & $C_1^{s}(\rho=0)$
       & $C_1^{s}(\rho=\rho_{\rm NM})$
       & $C_0^{\Delta s}$
       & $C_1^{\Delta s}$ 
\\ \hline
SIII   &   14.109   &    56.401 & 141.094 &  98.802  & 17.031 & 17.031 \\
SkI3   &   84.486   &  253.799  & 220.360 &  113.551 & 92.235 & 22.777 \\
SLy4   & --207.824  &  153.382  & 311.114 &  99.635  & 47.057 & 14.282 
\\ \hline
\end{tabular}
\end{center}
\end{table}
\clearpage
\begin{table}[htbp]
\caption{\label{p-o_SIIIt} 
Summary of the SIII-HFB calculation in keV. 
Notations; 
$\Omega^\pi_{\rm obl}$ and $\Omega^\pi_{\rm prol}$ 
denote the blocked level in the oblate and prolate minimum, respectively.
$\Delta E^{\rm HFBE}=E_{\rm obl}^{\rm HFBE}-E_{\rm prol}^{\rm HFBE}$,
$\Delta E^{\rm block}=E^{\rm block}_{\rm obl}-E^{\rm block}_{\rm prol}$, 
$\Delta E^{\rm pol}=E^{\rm pol}_{\rm obl}-E^{\rm pol}_{\rm prol}$, and
$\Delta E^{\rm HFB}=E_{\rm obl}^{\rm HFB}-E_{\rm prol}^{\rm HFB}$. 
Polarization contributions are further divided into
(i) the calculation with full time-odd terms, 
(ii) the calculation without $C^s_t$ terms, 
(iii) the calculation without $C^{\Delta s}_t$ terms, and 
(iv) the calculation without both  $C^s_t$ and $C^{\Delta s}_t$ terms. 
}
\begin{center}
 \begin{tabular}{c|c|c|c|c|c|c|c|c|c|c|c|}  \cline{2-12}
&$N$                 & 100   & 101     & 102   & 103  
                     & 104   & 105     & 106   & 107    
                     & 108   & 109  
\\ \cline{1-12}
\multicolumn{1}{|c|}{}
&$\Delta E^{\rm HFBE}$    &-- 167 & --110 & -- 68& -- 76
                          &-- 152 & --264 & --470& --807
                          &--1323 & --962
\\ \cline{2-12}
\multicolumn{1}{|c|}{}
&$\Omega^\pi_{\rm obl}$   & --    & 1/2$^-$ & --    & 1/2$^-$ 
                          & --    & 1/2$^-$ & --    & 1/2$^-$
                          & --    & 5/2$^-$
\\ \cline{2-12}
\multicolumn{1}{|c|}{}
&$\Omega^\pi_{\rm prol}$  & --    & 7/2$^-$ & --    & 7/2$^-$ 
                          & --    & 5/2$^-$ & --    & 9/2$^+$
                          & --    & 9/2$^-$
\\ \cline{2-12}
\multicolumn{1}{|c|}{}
&$E^{\rm block}_{\rm obl}$ 
                          & --    & 1220 & --    & 1263 
                          & --    & 1227 & --    & 1317
                          & --    & 1186
\\ \cline{2-12}
\multicolumn{1}{|c|}{}
 &$E^{\rm block}_{\rm prol}$
                          & --    &  934 & --    &  747
                          & --    &  854 & --    &  904
                          & --    & 1412
\\ \cline{2-12}
\multicolumn{1}{|c|}{}
&$\Delta E^{\rm block}$   & --    &  286 & --    &  516
                          & --    &  373 & --    &  413
                          & --    &--226
\\ \cline{1-12}
\multicolumn{1}{|c|}{(i)}
&$E^{\rm pol}_{\rm obl}$ 
                          & --    &-- 34 & --    &-- 39
                          & --    &-- 10 & --    &-- 10
                          & --    &    7
\\ \cline{2-12}
\multicolumn{1}{|c|}{}
 &$E^{\rm pol}_{\rm prol}$
                          & --    &   32 & --    &   31
                          & --    &   36 & --    &   47
                          & --    &   36
\\ \cline{2-12}
\multicolumn{1}{|c|}{}
&$\Delta E^{\rm pol}$   & --    &-- 65 & --    &-- 70
                        & --    &-- 46 & --    &-- 58
                        & --    &-- 30
\\ \cline{2-12}
\multicolumn{1}{|c|}{}
&$\Delta E^{\rm HFB}$   &-- 167 &   110 &-- 68 &  370
                        &-- 152 &    63 &--470 &--452
                        &--1323 &--1217
\\ \hline
\multicolumn{1}{|c|}{(ii)}
&$E^{\rm pol}_{\rm obl}$ 
                     & --    &--317 & --    &--328 
                     & --    &--246 & --    &--250
                     & --    &--173 
\\ \cline{2-12}
\multicolumn{1}{|c|}{}
&$E^{\rm pol}_{\rm prol}$
                        & --    &--141 & --    &--102
                        & --    &--142 & --    &--111
                        & --    &--147
\\ \cline{2-12}
\multicolumn{1}{|c|}{}
&$\Delta E^{\rm pol}$   & --    &--176 & --    &--226
                        & --    &--104 & --    &--139
                        & --    &-- 26 
\\ \cline{2-12}
\multicolumn{1}{|c|}{}
&$\Delta E^{\rm HFB}$   & --167 &--   1 & -- 68 &  214
                        & --152 &     5 & --470 &--533
                        & --1323&--1214
\\ \hline
\multicolumn{1}{|c|}{(iii)}
&$E^{\rm pol}_{\rm obl}$ 
                     & --    &   32 & --    &   58
                     & --    &   79 & --    &   76
                     & --    &   49
\\ \cline{2-12}
\multicolumn{1}{|c|}{}
&$E^{\rm pol}_{\rm prol}$
                     & --    &   60 & --    &   58
                     & --    &   73 & --    &   73
                     & --    &   61 
\\ \cline{2-12}
\multicolumn{1}{|c|}{}
&$\Delta E^{\rm pol}$   & --    &-- 28 & --    &   0
                        & --    &    6 & --    &   3
                        & --    &-- 11
\\ \cline{2-12}
\multicolumn{1}{|c|}{}
&$\Delta E^{\rm HFB}$   & -- 167&   147 & -- 68 &  440
                        & -- 152&   115 & -- 470&--391
                        & --1323&--1199
\\ \hline
\multicolumn{1}{|c|}{(iv)}
&$E^{\rm pol}_{\rm obl}$ 
                     & --    &-- 31 & --    &-- 32
                     & --    &-- 20 & --    &-- 16
                     & --    &-- 43
\\ \cline{2-12}
\multicolumn{1}{|c|}{}
&$E^{\rm pol}_{\rm prol}$
                     & --    &-- 66 & --    &-- 64
                     & --    &-- 41 & --    &-- 41
                     & --    &-- 80
\\ \cline{2-12}
\multicolumn{1}{|c|}{}
&$\Delta E^{\rm pol}$   & --    &   35 & --    &   32
                        & --    &   22 & --    &   25
                        & --    &   36
\\ \cline{2-12}
\multicolumn{1}{|c|}{}
&$\Delta E^{\rm HFB}$   & -- 167&   210 & -- 68 &   472
                        & -- 152&   131 & -- 470& --370
                        & --1323&--1152
\\ \hline
 \end{tabular}
\end{center}
\end{table}

\clearpage
\begin{table}[htbp]
\caption{\label{p-o_SkI3t} 
Summary of the SkI3-HFB calculation in keV. 
Notations are the same as those in Table 2. 
}
\begin{center}
 \begin{tabular}{c|c|c|c|c|c|c|c|c|c|c|c|}  \cline{2-12}
&$N$                 & 100    & 101      & 102    & 103  
                     & 104    & 105      & 106    & 107     
                     & 108    & 109  
\\ \cline{1-12}
\multicolumn{1}{|c|}{}
&$\Delta E^{\rm HFBE}$  &--  54 &    20 &-- 93 &-- 90
                        &--  76 &-- 106 &--268 &--628
                        &--1159 &--1855
\\ \cline{2-12}
\multicolumn{1}{|c|}{}
&$\Omega^\pi_{\rm obl}$   & --    & 1/2$^-$ & --    & 1/2$^-$ 
                          & --    & 1/2$^-$ & --    & 5/2$^-$
                          & --    & 3/2$^-$
\\ \cline{2-12}
\multicolumn{1}{|c|}{}
&$\Omega^\pi_{\rm prol}$  & --    & 7/2$^-$ & --    & 7/2$^-$ 
                          & --    & 5/2$^-$ & --    & 9/2$^+$
                          & --    & 3/2$^-$
\\ \cline{2-12}
\multicolumn{1}{|c|}{}
&$E^{\rm block}_{\rm obl}$ 
                     & --    & 1017 & --    & 1124
                     & --    & 1362 & --    & 1098
                     & --    & 1072
\\ \cline{2-12}
\multicolumn{1}{|c|}{}
 &$E^{\rm block}_{\rm prol}$
                     & --    &  865 & --    &  691
                     & --    &  828 & --    &  887
                     & --    &  885
\\ \cline{2-12}
\multicolumn{1}{|c|}{}
&$\Delta E^{\rm block}$ & --    &  152 & --    &  433
                        & --    &  534 & --    &  211
                        & --    &  187 
\\ \cline{1-12}
\multicolumn{1}{|c|}{(iii)}
&$E^{\rm pol}_{\rm obl}$ 
                     & --    &  124 & --    &  122 
                     & --    &  116 & --    &   73
                     & --    &   87 
\\ \cline{2-12}
\multicolumn{1}{|c|}{}
 &$E^{\rm pol}_{\rm prol}$
                     & --    &   41 & --    &   35 
                     & --    &  105 & --    &   53 
                     & --    &   91 
\\ \cline{2-12}
\multicolumn{1}{|c|}{}
&$\Delta E^{\rm pol}$& --    &   84 & --    &   87
                     & --    &   11 & --    &   20
                     & --    &--  4
\\ \cline{2-12}
\multicolumn{1}{|c|}{}
&$\Delta E^{\rm HFB}$&--  54 &   256 &-- 93 &  430
                     &--  76 &   440 &--268 &--397
                     &--1159 &--1673
\\ \hline
\multicolumn{1}{|c|}{(iv)}
&$E^{\rm pol}_{\rm obl}$ 
                     & --    &-- 27 & --    &-- 26
                     & --    &-- 29 & --    &-- 73
                     & --    &-- 59
\\ \cline{2-12}
\multicolumn{1}{|c|}{}
&$E^{\rm pol}_{\rm prol}$
                     & --    &--194 & --    &--193
                     & --    &--102 & --    &--147
                     & --    &--126
\\ \cline{2-12}
\multicolumn{1}{|c|}{}
&$\Delta E^{\rm pol}$& --    &   168 & --    &  167
                     & --    &    73 & --    &   74
                     & --    &    67
\\ \cline{2-12}
\multicolumn{1}{|c|}{}
&$\Delta E^{\rm HFB}$& -- 54 &   340 & -- 93 &  509
                     & -- 76 &   501 & --268 &--343
                     & --1159&--1602
\\ \hline
 \end{tabular}
\end{center}
\end{table}

\clearpage
\begin{table}[htbp]
\caption{\label{p-o_SLy4t} 
Summary of the SLy4-HFB calculation in keV. 
Notations are the same as those in Table 2.
}
\begin{center}
 \begin{tabular}{c|c|c|c|c|c|c|c|c|c|c|c|}  \cline{2-12}
&$N$                 & 100    & 101      & 102    & 103  
                     & 104    & 105      & 106    & 107     
                     & 108    & 109  
\\ \cline{1-12}
\multicolumn{1}{|c|}{}
&$\Delta E^{\rm HFBE}$&-- 181 &-- 219 &-- 276 &-- 428
                      &-- 670 &-- 968 &--1337 &--1457
                      &--1530 &--1601 
\\ \cline{2-12}
\multicolumn{1}{|c|}{}
&$\Omega^\pi_{\rm obl}$   & --    & 1/2$^-$ & --    & 1/2$^-$ 
                          & --    & 1/2$^-$ & --    & 3/2$^-$
                          & --    & 3/2$^-$
\\ \cline{2-12}
\multicolumn{1}{|c|}{}
&$\Omega^\pi_{\rm prol}$  & --    & 5/2$^-$ & --    & 5/2$^-$ 
                          & --    & 5/2$^-$ & --    & 1/2$^-$
                          & --    & 1/2$^-$
\\ \cline{2-12}
\multicolumn{1}{|c|}{}
&$E^{\rm block}_{\rm obl}$ 
                     & --    & 1127 & --    & 1109
                     & --    & 1271 & --    & 1169
                     & --    & 1080
\\ \cline{2-12}
\multicolumn{1}{|c|}{}
 &$E^{\rm block}_{\rm prol}$
                     & --    &  849 & --    &  896
                     & --    &  810 & --    & 1276
                     & --    & 1127
\\ \cline{2-12}
\multicolumn{1}{|c|}{}
&$\Delta E^{\rm block}$ & --    &  277 & --    &  212
                        & --    &  461 & --    &--107
                        & --    &-- 47
\\ \cline{1-12}
\multicolumn{1}{|c|}{(i)}
&$E^{\rm pol}_{\rm obl}$ 
                     & --    &-- 82 & --    &-- 68
                     & --    &-- 69 & --    &-- 16
                     & --    &-- 14
\\ \cline{2-12}
\multicolumn{1}{|c|}{}
 &$E^{\rm pol}_{\rm prol}$
                     & --    &   19 & --    &-- 17
                     & --    &   13 & --    &-- 30
                     & --    &-- 40

\\ \cline{2-12}
\multicolumn{1}{|c|}{}
&$\Delta E^{\rm pol}$& --    &--101 & --    &-- 50
                     & --    &-- 81 & --    &   15
                     & --    &   26
\\ \cline{2-12}
\multicolumn{1}{|c|}{}
&$\Delta E^{\rm HFB}$&-- 181 &--  42 &-- 276 &-- 266
                     &-- 670 &-- 588 &--1337 &--1550
                     &--1530 &--1622
\\ \hline
\multicolumn{1}{|c|}{(iii)}
&$E^{\rm pol}_{\rm obl}$ 
                     & --    &  127 & --    &  117
                     & --    &  131 & --    &   91
                     & --    &   85
\\ \cline{2-12}
\multicolumn{1}{|c|}{}
&$E^{\rm pol}_{\rm prol}$
                     & --    &  121 & --    &   44
                     & --    &   49 & --    &  140
                     & --    &  139
\\ \cline{2-12}
\multicolumn{1}{|c|}{}
&$\Delta E^{\rm pol}$& --    &     7 & --    &   74
                     & --    &    82 & --    &-- 49
                     & --    &--  55 
\\ \cline{2-12}
\multicolumn{1}{|c|}{}
&$\Delta E^{\rm HFB}$& -- 181&    65 & -- 276&-- 142
                     & -- 670&-- 425 & --1337&--1631
                     & --1530&--1702
\\ \hline
\multicolumn{1}{|c|}{(iv)}
&$E^{\rm pol}_{\rm obl}$ 
                     & --    &-- 14 & --    &-- 22
                     & --    &-- 22 & --    &-- 39
                     & --    &-- 36 
\\ \cline{2-12}
\multicolumn{1}{|c|}{}
&$E^{\rm pol}_{\rm prol}$
                     & --    &-- 66 & --    &-- 101
                     & --    &--116 & --    &--  13
                     & --    &-- 13
\\ \cline{2-12}
\multicolumn{1}{|c|}{}
&$\Delta E^{\rm pol}$& --    &    52 & --    &   79 
                     & --    &    94 & --    &-- 25
                     & --    &--  22
\\ \cline{2-12}
\multicolumn{1}{|c|}{}
&$\Delta E^{\rm HFB}$& -- 181&   110 & -- 276&-- 137
                     & -- 670&-- 413 & --1337&--1589
                     & --1530&--1670
\\ \hline
 \end{tabular}
\end{center}
\end{table}
\clearpage
%
%
\end{document}